\documentclass[twocolumn,showpacs,preprintnumbers]{revtex4-1}
\usepackage{graphicx,color}
\usepackage{bm}
\begin{document}

\title{Random-matrix approach to the statistical compound nuclear
  reaction at low energies using the Monte-Carlo technique}

\author{T. Kawano}
\email{kawano@lanl.gov}
\affiliation{Theoretical Division, Los Alamos National Laboratory, Los Alamos, NM 87545, USA}

\author{P. Talou}
\affiliation{Theoretical Division, Los Alamos National Laboratory, Los Alamos, NM 87545, USA}

\author{H. A. Weidenm\"{u}ller}
\affiliation{Max-Planck-Institut f\"{u}r Kernphysik, 69029 Heidelberg, Germany}

\date{\today}

\begin{abstract}
Using a random-matrix approach and Monte-Carlo simulations, we
generate scattering matrices and cross sections for compound-nucleus
reactions. In the absence of direct reactions we compare the average
cross sections with the analytic solution given by the Gaussian
Orthogonal Ensemble (GOE) triple integral, and with predictions of
statistical approaches such as the ones due to Moldauer, to Hofmann,
Richert, Tepel, and Weidenm\"{u}ller, and to Kawai, Kerman, and
McVoy. We find perfect agreement with the GOE triple integral and
display the limits of validity of the latter approaches. We establish
a criterion for the width of the energy-averaging interval such that
the relative difference between the ensemble-averaged and the
energy-averaged scattering matrices lies below a given bound. Direct
reactions are simulated in terms of an energy-independent background
matrix. In that case, cross sections averaged over the ensemble of
Monte-Carlo simulations fully agree with results from the
Engelbrecht-Weidenm\"{u}ller transformation. The limits of other
approximate approaches are displayed.
\end{abstract}
\maketitle

\section{Introduction}
\label{sec:introduction}

For medium-weight and heavy target nuclei, nuclear reactions
represent a very complex phenomenon because the number of degrees of
freedom grows rapidly with mass number $A$. That fact has naturally
led to the development of a statistical approach. Central to the
approach are the concept of a fully equilibrated compound nucleus and
the Bohr hypothesis \cite{Bohr36}, which states that a
particle incident on a medium-weight or heavy nucleus shares its
energy with the target nucleons. The resulting compound nucleus
attains statistical equilibrium, and the modes of decay of the
equilibrated system are independent of the mode of formation. The
postulated independence implies a factorization of the energy-averaged
compound-nucleus cross section \cite{Hauser52}. The factorization
hypothesis holds very well at sufficiently large bombarding energies
(i.e., in the Ericson regime) but not for isolated or weakly
overlapping compound-nucleus resonances \cite{Lane57,Moldauer75a}. In
that regime the average cross section given by the factorization
hypothesis must be corrected by a ``width fluctuation correction''
factor (WFC). The WFC factor basically accounts for an enhancement of
the elastic average cross section.

In the 1970s numerous efforts were undertaken to derive the WFC
factor \cite{Moldauer75a, Moldauer75b, Kawai73, Agassi75, Hofmann75,
Mello79, Mello80} or to generate a suitable parametrization of the WFC
factor with the help of the Monte-Carlo (MC) technique
\cite{Hofmann75, Hofmann80, Moldauer78, Moldauer80}. All of these
were guided by random-matrix theory (RMT). Inspired by Bohr's idea,
Wigner had introduced RMT into nuclear physics as a means to cope with
the complexities of the compound nucleus (see Brody {\it et al.}
\cite{Brody81}). In RMT, the nuclear Hamiltonian is assumed to be a
member of the Gaussian Orthogonal Ensemble (GOE) of random matrices.
Wigner himself never went as far as formulating a statistical theory
of nuclear reactions in terms of the GOE. Lacking such a theory, the
above-mentioned approaches used approximations that were not fully
controlled. Only in 1985 an exact closed-form expression for the
average $S$ matrix and for the $S$ matrix correlation function based
upon a GOE scattering approach was derived \cite{Verbaarschot85},
based upon the shell-model approach to nuclear reactions
\cite{Mahaux69} and valid in the limit of a large number of resonances.
In that work the $S$ matrix is written in terms of the GOE Hamiltonian
$H^{({\rm GOE})}$. Averages are performed directly over the
Gaussian-distributed elements of $H^{({\rm GOE})}$.

The exact results of the GOE scattering
approach \cite{Verbaarschot85} apply for all values of the parameters
(number of open channels, isolated or overlapping resonances)
characterizing compound-nucleus reactions. More generally, that work
describes universal features of quantum-chaotic scattering
\cite{Weidenmueller09, Mitchel10} and is, therefore, relevant also
beyond the confines of nuclear physics. However, the exact expression
for the $S$-matrix correlation function \cite{Verbaarschot85} involves
a triple integral. The computational cost of evaluating that integral
is quite heavy especially when many channels are open. That is why
only few numerical studies have been performed in the past.
Fr\"{o}hner \cite{Frohner89} and Igarasi \cite{Igarasi91}
independently compared the GOE triple integral results with Moldauer's
method, and obtained good agreement. Hilaire, Lagrange, and
Koning \cite{Hilaire03} extended the numerical study, and applied it
to some realistic cases where neutron radiative capture and fission
channels are involved. Updated parametrizations of Moldauer's method
based on the GOE triple integral calculation are also available
\cite{Ernebjerg04, Kawano13}, which are of practical use for
cross-section calculations.

In the present paper we present a thorough analysis of the
results of the GOE approach and a comparison with other, approximate
methods, with the aim to understand their applicability and
limitation. The work is based upon a Monte-Carlo approach. We generate
an ensemble of scattering matrices or cross sections. This is done by
drawing at random the elements of $H^{({\rm GOE})}$ and using these to
generate the elements of the scattering matrix. In this way we are
able to avoid some phenomenological assumptions made in the past
concerning the distribution of the decay amplitudes or of levels. In
that respect our MC approach also differs from the one used by
Moldauer or Hofmann {\it et al.} We average over the ensemble of
realizations generated by the MC method and compare these with
predictions of the exact GOE approach and of other approximate
methods. We are able to answer some important and long-standing
questions concerning compound nuclear reactions, such as the
difference between energy and ensemble averages, the role of direct
channels, the existence of correlations between distributions of levels
and decay amplitudes, and the behavior of the cross section in the limit
of weak absorption.

\section{Theory of stochastic scattering}
\label{sec:theory}

  \subsection{Compound-nucleus cross section}
  \label{subsec:theoryCN}

The cross section for a reaction from channel $a$ to channel $b$ is
written as
\begin{equation}
   \sigma_{ab} = \frac{\pi}{k_a^2} g_a | \delta_{ab} - S_{ab} |^2 \ .
\end{equation}
Here $k_a$ is the wave number for channel $a$, $g_a$ is the spin
factor, and the element $S_{ab}$ of the scattering matrix $S$ consists
of an energy-averaged part $\langle S_{ab}\rangle$ and a fluctuating
part $S_{ab}^{{\rm fl}}$. The energy-averaged cross section also
consists of two parts,
\begin{eqnarray}
   \langle\sigma_{ab}\rangle
   & = & \frac{\pi}{k_a^2} g_a \langle | \delta_{ab} - S_{ab}|^2 \rangle
   \nonumber\\
   & = & \frac{\pi}{k_a^2} g_a
         \left\{
           | \delta_{ab} - \langle S_{ab} \rangle|^2
           + \langle | S_{ab}^{{\rm fl}} |^2 \rangle
         \right\} \ .
   \label{eq:sigma}
\end{eqnarray}
The term containing $|\delta_{ab} - \langle S_{ab} \rangle|^2$
describes shape elastic ($a = b$) or shape inelastic ($a \neq b$)
scattering. The term containing $\langle | S_{ab}^{{\rm fl}} |^2
\rangle$ is the average compound-nucleus (CN) cross section
\begin{equation}
  \sigma^{\rm CN}_{a b}
  = \frac{\pi}{k_a^2} g_a \langle |S^{\rm fl}_{ab}|^2 \rangle \ .
\end{equation}
In the first part of the paper we confine ourselves to
cases where the average $S$ matrix is diagonal, $\langle S_{a b}
\rangle = \delta_{a b} \langle S_{a a} \rangle$. Then $\langle S_{a a}
\rangle$ and the shape-elastic cross section
\begin{equation}
  \sigma^{\rm SE}_{a a}
  = \frac{\pi}{k_a^2} g_a | 1 - \langle S_{aa} \rangle|^2
\end{equation}
are given by the optical model. It is the aim of various theories of
CN reactions to express the CN cross section in terms of $\langle S_{a
  a} \rangle$ and of the transmission coefficients
\begin{equation}
  T_a = 1 - | \langle S_{aa} \rangle|^2 \ , \
  \quad 0 \le T_a \le 1 \ .
  \label{eq:transmission}
\end{equation}
These measure the unitary deficit of $\langle S \rangle$ and, thus,
the probability of CN formation. In the second part of the paper we
address the case when $\langle S_{a b} \rangle$ is not diagonal.

Bohr's idea of the independence of formation and decay of the CN led
to the Hauser-Feshbach formula~\cite{Hauser52} for the CN cross section,
\begin{equation}
   \sigma^{\rm HF}_{ab}
   = \frac{\pi}{k_a^2} g_a \frac{T_a T_b}{\sum_c T_c} \ .
   \label{eq:HF}
\end{equation}
Corrections to that formula are conveniently expressed in terms of
the ``width fluctuation correction'' (WFC) factor~\cite{Moldauer61},
\begin{equation}
   \sigma^{\rm CN}_{ab}
   = \frac{\pi}{k_a^2} g_a \frac{T_a T_b}{\sum_c T_c} W_{ab} \ .
   \label{eq:WFC}
\end{equation}
Rigorously speaking, $W_{ab}$ should be separated into two parts, the
``elastic enhancement factor'' and the proper ``width fluctuation
correction factor''~\cite{Moldauer75a}. However, for the comparison of
various approaches it is more convenient to adopt the suggestion of
Hilaire, Lagrange, and Koning~\cite{Hilaire03} and to define the width
fluctuation factor $W_{ab}$ as the ratio $\sigma^{\rm CN}_{ab} /
\sigma^{\rm HF}_{ab}$.

In what follows we compare several approaches to the calculation of
$\sigma^{\rm CN}_{ab}$ and/or of the WFC factor. In chronological
order, these are the approach of Kawai, Kerman, and
MacVoy~\cite{Kawai73} (KKM), the parametrization by Hofmann, Richert,
Tepel, and Weidenm\"{u}ller~\cite{Hofmann75} (HRTW), Moldauer's
parametrization~\cite{Moldauer80}, the GOE approach by Verbaarschot,
Weidenm\"{u}ller, and Zirnbauer~\cite{Verbaarschot85}, the
parametrization by Ernebjerg and Herman~\cite{Ernebjerg04}, and that
by Kawano and Talou~\cite{Kawano13}. These are briefly summarized in
the Appendix. Hereafter we drop the kinetic and spin factors $\pi g_a
/ k_a^2$, so that all the cross sections are dimensionless.

All these approaches use GOE-inspired statistical assumptions on the
parameters of the CN resonances. In our comparison we use the results
of Ref.~\cite{Verbaarschot85} as a benchmark. We do so because the
work of Ref.~\cite{Verbaarschot85} is the only one that, starting from
a random-matrix model for the Hamiltonian of the CN resonances and
using controlled approximations, obtains an analytical expression for
$\sigma^{\rm CN}_{ab}$ that is valid in all regimes ---
from the regime of isolated resonances to that of strongly overlapping
resonances.

  \subsection{$S$ matrix, $K$ matrix, $R$ matrix}
  \label{subsec:theoryS}

In order to display the connection between various theories of
resonance reactions we recall here briefly the derivation of a
universal expression for the $S$ matrix~\cite{Mahaux69, Kawai73}.
Specialization of that expression then yields the formulas used in
various approaches.

Given a time-reversal-invariant Hamiltonian $H$ we use Feshbach's
projection operators $P$ and $Q = 1 - P$ (where $P$ projects onto all
open channels labelled $a, b, \ldots$) to write the Schr\"{o}dinger
equation $(E-H) \Psi = 0$ for the scattering wave function $\Psi$ in
the form of the coupled equations
\begin{eqnarray}
  (E - H_{PP}) P \Psi &=& H_{PQ} Q \Psi \ , \\
  (E - H_{QQ}) Q \Psi &=& H_{QP} P \Psi \ .
\end{eqnarray}
We use the standard notation, $H_{PP} = PHP$, $H_{PQ} = PHQ$, etc. With
the $P$ space scattering wave function $\psi_a^{(+)}$ defined by
\begin{equation}
  (E - H_{PP}) \psi_a^{(+)} = 0 \ ,
\end{equation}
the unitary and symmetric $S$ matrix is given by
\begin{equation}
  S_{ab} = S_{ab}^{(0)} - 2\pi i
          \left(
          \psi_a^{(-)}
             | H_{PQ} \frac{1}{E - {\cal H}_{QQ}} H_{QP} |
          \psi_b^{(+)} 
          \right) \ .
\label{eq:Smatrix}
\end{equation}
Here $S_{ab}^{(0)}$ is a unitary background scattering matrix defined
by the asymptotic form of the solutions $\psi_a^{(+)}$, and ${\cal
H}_{QQ}$ is the effective Hamiltonian in $Q$ space,
\begin{equation}
  {\cal H}_{QQ} = H_{QQ} + H_{QP} \frac{1}{E^+ - H_{PP}} H_{PQ} \ .
\label{eq:Heff}
\end{equation}
To be useful Eqs.~(\ref{eq:Smatrix}) and (\ref{eq:Heff}) must be
specialized further. The $S$-matrix approach of
Ref.~\cite{Verbaarschot85} and the expressions for $S$ in terms of the
$K$ matrix and the $R$ matrix use different such specializations.
Common to these is the assumption that the unitary background
scattering matrix $S^{(0)}$ is diagonal, $S^{(0)}_{a b} = \delta_{a
b} \exp \{ 2 i \phi_a \}$. We assume that the phases $\phi_a$ are
removed by the transformation $S_{a b} \to \exp \{ - i \phi_a \} S_{a
b} \exp \{ - i \phi_b \}$. Then $\psi^{(+)}_a \exp \{ - i \phi_a \}
= \psi_a$ is real, and Eq.~(\ref{eq:Smatrix}) becomes
\begin{equation}
  S_{ab} = \delta_{ab} - 2\pi i
          \left(
          \psi_a
             | H_{PQ} \frac{1}{E - {\cal H}_{QQ}} H_{QP} |
          \psi_b 
          \right) \ .
\label{eq:Smatrix1}
\end{equation}

For the $S$-matrix approach of Ref.~\cite{Verbaarschot85} we introduce
an arbitrary orthonormal basis of states labeled $\mu$ in $Q$ space
and write
\begin{equation}
  W_{\mu a} = ( \mu | H_{Q P} | \psi_a ) = W_{a \mu} = W^*_{a \mu} \ ,
 \label{eq:VWZ1}
\end{equation}
\begin{equation}
  \left(
     \mu | H_{QP} \frac{1}{E^+ - H_{PP}} H_{PQ} | \nu
   \right)
  = \Delta_{\mu \nu} - i \pi \sum_c W_{\mu c} W_{c \nu} \ ,
 \label{eq:VWZ2}
\end{equation}
\begin{equation}
 \left( \mu | H_{Q Q} | \nu \right) = H_{\mu \nu}  \ .
 \label{eq:VWZ3}
\end{equation}
The sum extends over all open channels. The real shift function
$\Delta_{\mu \nu}$ is defined by a principal-value integral. It is
commonly assumed that the matrix elements $W_{a \mu}$ change slowly
with energy on a scale defined by the mean level spacing of the
resonances. Then $\Delta_{\mu \nu} \approx 0$. We use that assumption
throughout. With these definitions, Eqs.~(\ref{eq:Smatrix}) and
(\ref{eq:Heff}) take the form
\begin{equation}
  S_{a b} = \delta_{a b}
         - 2 i \pi \sum_{\mu \nu} W_{a \mu} (D^{- 1})_{\mu\nu} W_{\nu b} 
\label{eq:Sreal}
\end{equation}
where
\begin{equation}
  D_{\mu \nu} = E \delta_{\mu \nu}
            - H_{\mu \nu} + i \pi \sum_{c} W_{\mu c}W_{c \nu } \ .
\label{prop}
\end{equation}

For the $K$-matrix parametrization of $S$ we use Eq.~(\ref{eq:Heff})
to define the eigenvalues $E_\sigma$ and eigenvectors $X_\sigma$ of
the bound compound system,
\begin{equation}
  (E_\sigma - H_{QQ}) X_\sigma = 0 \ .
\end{equation}
The states $X_\sigma$ produce the CN resonances in the scattering
process. These states correspond to a special choice of the basis of
states $\mu$ used in Eqs.~(\ref{eq:VWZ1})--(\ref{eq:VWZ3}). The
partial decay amplitude of state $X_\sigma$ into channel $a$ is
\begin{equation}
  \gamma_{\sigma a} = \gamma^*_{\sigma a}
   = ( X_\sigma | H_{QP} | \psi_a ) \ .
  \label{eq:amplitude}
\end{equation}
Under neglect of the shift matrix $\Delta$ the $S$ matrix of
Eq.~(\ref{eq:Smatrix1}) can be written as
\begin{equation}
  S_{a b} = \bigg( \frac{1 - i K}{1 + i K} \bigg)_{a b}
  \label{eq:SKmatrix}
\end{equation}
where
\begin{equation}
  K_{a b}(E)
   = \pi \sum_\sigma \frac{\gamma_{a \sigma} \gamma_{\sigma b}}{E - E_\sigma}
   \ .
  \label{eq:Kmatrix}
\end{equation}

The $R$ matrix is obtained by a non-standard choice of the projection
operators $P$ and $Q$. In every channel $c$ (open and closed) a radius
$r_c$ is defined. The set $\{ r_c \}$ of all channel radii separates
the internal and the external regions of configuration space. The
operator $Q$ projects onto the internal region, and $P = 1 - Q$. At
the channel surfaces of the internal region, self-adjoint boundary
conditions are introduced with real boundary condition parameters
$B_c$. These define a Hermitian Hamiltonian $H_{Q Q}$ and associated
internal eigenvalues $E^{({\rm R})}_\sigma$ and orthonormal eigenfunctions
$X^{({\rm R})}_\sigma$. The reduced width amplitude
$\gamma^{({\rm R})}_{\sigma c}$ is the projection of the eigenfunction
$X^{({\rm R})}_\sigma$ onto the surface of channel $c$, and the $R$ matrix is
defined as
\begin{equation}
  R_{a b}(E)
   = \sum_\sigma
     \frac{\gamma^{({\rm R})}_{a \sigma} \gamma^{({\rm R})}_{\sigma b}}
          {E^{({\rm R})}_\sigma - E} \ .
  \label{eq:Rmatrix}
\end{equation}
This is in close analogy to Eq.~(\ref{eq:Kmatrix}), except that a
factor $\sqrt{\pi}$ has been absorbed by each of the reduced width
amplitudes. The resulting form of the scattering matrix is
\begin{eqnarray}
  S_{a b} &=& \delta_{a b} \nonumber\\
         &+& 2 i
            \sqrt{P_a}
                \left(
                  \{1 - R (L - B)\}^{- 1} R
                \right)_{a b}
            \sqrt{P_b} \ .
\label{SRmatrix}
\end{eqnarray}
Here $P_a$ is the penetration factor in channel $a$, the matrices $B$
and $L$ are diagonal with elements $B_a$ and $L_a = {\cal S}_a + i
P_a$, and the real entities ${\cal S}_a - B_a$ play a role that is
analogous to that of the shift function $\Delta$ in
Eq.~(\ref{eq:VWZ2}). The diagonal matrices $L$ and
$P$ depend only on channel radius $r_a$ and wave number
$k_a$~\cite{Moldauer64}. The boundary condition parameter $B_a$ is
often taken as $B_a = -l_a$~\cite{Wigner47}, with $l_a$ the orbital
angular momentum of relative motion in channel $a$. In the $R$-matrix
approach the phases $\phi_a$ are caused by elastic scattering on a
hard sphere of radius $r_a$ while in the approach of
Ref.~\cite{Verbaarschot85} they are elastic potential scattering phase
shifts.

  \subsection{Implementation of stochasticity}
  \label{subsec:theoryGOE}

To fully define the scattering matrices in Sec.~\ref{subsec:theoryS}
we need to determine the resonance parameters. This is done by
introducing statistical assumptions, using random-matrix
theory~\cite{Mehta04} as a guiding principle. The actual procedure is
somewhat different for the three forms of the scattering matrix in
Sec.~\ref{subsec:theoryS}. These are referred to as the $S$-matrix
approach, the $K$-matrix approach, and the $R$-matrix approach,
respectively.

The relevant random-matrix ensemble is the time-reversal-invariant
Gaussian Orthogonal Ensemble (GOE). The elements $H^{({\rm GOE})}_{\mu
  \nu}$ of the $N$-dimensional GOE matrix $H^{({\rm GOE})}$ are
Gaussian-distributed real random variables with zero mean values and
second moments given by
\begin{equation}
 \overline{ H_{\mu\nu}^{({\rm GOE})} H_{\rho\sigma}^{({\rm GOE})} }
 = \frac{\lambda^2}{N}
   (\delta_{\mu\rho}\delta_{\nu\sigma} + \delta_{\mu\sigma}\delta_{\nu\rho}) \ .
 \label{eq:GOE}
\end{equation}
Here and in what follows, the ensemble average is denoted by an
overbar. The parameter $\lambda$ is related to the average level
spacing $d$ at the center of the GOE spectrum by $d = \pi \lambda /
N$. Universal properties are analytically derived~\cite{Mehta04} in
the limit $N \to \infty$. These are: The eigenvalues and the
eigenvectors of $H^{({\rm GOE})}$ are statistically uncorrelated. The
projections of the (real) eigenvectors on any fixed vector in Hilbert
space have a Gaussian distribution centered at zero. The eigenvalues
obey Wigner-Dyson statistics. The degree to which these properties can
be implemented depends on the approach used.

In the $S$-matrix approach the $Q$-space Hamiltonian $H_{\mu \nu}$ of
Eq.~(\ref{prop}) is replaced by $H^{({\rm GOE})}_{\mu \nu}$. That
replacement provides the most direct implementation of random-matrix
theory into scattering theory. The average cross section is worked out
as an average over the GOE. For the specification of the parameters
$W_{a \mu}$ one uses the invariance of the GOE under orthogonal
transformations in Hilbert space. That invariance implies that
ensemble averages can depend on the $W$'s only via the invariant forms
$\sum_\mu W_{a \mu} W_{\mu b}$. For the average $S$ matrix to be
diagonal, the sums must be diagonal in the channel indices,
\begin{equation}
   \sum_\mu W_{a \mu} W_{\mu b} = \delta_{a b} N v^2_a \ ,
\label{eq:mel}
\end{equation}
and the only parameters left are the $v^2_a$. With
\begin{equation}
   x_a = \frac{\pi^2 v^2_a}{d} \ ,
   \label{eq:xa}
\end{equation}
these determine the average $S$ matrix elements $\overline{S}_{aa}$
and the transmission coefficients $T_a$ as
\begin{equation}
  \overline{S}_{aa} = \frac{1 - x_a}{1 + x_a} \ , \qquad
  T_a = \frac{4 x_a}{(1 + x_a)^2}  \ .
  \label{eq:trans}
\end{equation}
Equations~(\ref{eq:trans}) imply that the average
strength $x_a$ of the coupling of the CN resonance states to channel
$a$ is fixed by the average $S$ matrix and, thus, determined by the
shape-elastic input. In that sense, the GOE ensemble average of
$|S^{\rm fl}_{a b}|^2$ is parameter-free. This is different from past
calculations using a statistical $R$ matrix or $K$ matrix.

In the $K$-matrix and $R$-matrix approaches, two
assumptions are made:
\begin{enumerate}
   \item The partial width amplitudes $\gamma_{a \sigma}$ and the
    reduced width amplitudes $\gamma^{({\rm R})}_{a \sigma}$ both have
    a Gaussian distribution with zero mean and a specified second
    moment $\langle \gamma_a^2 \rangle$.

   \item The eigenvalues $E_\sigma$ and $E^{({\rm R})}_\sigma$
    obey Wigner-Dyson statistics.
\end{enumerate}
If fully implemented, these assumptions correspond for $N \to \infty$
to the properties of the GOE listed below Eq.~(\ref{eq:GOE}).

In practical calculations, the implementation of these statistical
assumptions causes difficulties. The $S$-matrix approach lends itself
to an analytical calculation of $\sigma^{{\rm CN}}_{ab}$ in the limit
$N \to \infty$. The resulting expression is given in
Eq.~(\ref{eq:GOE3int}) below. However, the use of that expression was
limited for a long time because of the difficulties in calculating
reliably the ensuing threefold integral. A direct implementation would
consist in drawing the elements $H^{({\rm GOE})}_{\mu \nu}$ from a
Gaussian distribution, choosing a set of matrix elements $W_{a \mu}$
consistent with Eqs.~(\ref{eq:mel}) and (\ref{eq:trans}), and
inverting the resulting matrix $D_{\mu \nu}$. For $N \gg 1$ that is
quite cumbersome and, to the best of our knowledge, has not been done
before. We report on such a calculation below.

For the $K$-matrix and $R$-matrix approaches, it is straightforward to
draw the partial width amplitudes or the reduced width amplitudes from
a Gaussian distribution. To meet postulate 2, the eigenvalues should
be determined by diagonalization of the GOE matrix $H^{({\rm GOE})}$
for $N \gg 1$. This is cumbersome, and a simplified version sometimes
replaces postulate 2. The Wigner surmise for the distribution $P(s)$
of spacings $s$ of neighboring eigenvalues reads
\begin{equation}
  P_W(s) = \frac{\pi}{2} s \exp\left(-\frac{\pi s^2}{4}\right) \ ,
  \label{eq:wigner}
\end{equation}
with $s$ the actual spacing in units of $d$. Spacings of neighboring
eigenvalues are drawn at random from $P_W(s)$ and are used to
construct the spectrum. Higher correlations between eigenvalue
spacings are thereby neglected. In particular, the stiffness of the
GOE spectrum (a central property) is not taken into account.

With this input, the energy-averaged cross section $\langle
|\delta_{ab} - S_{ab}|^2 \rangle$ can be calculated. It is often
assumed that the energy average can be replaced by an ensemble average
over the joint distribution of level energies and decay amplitudes.
The ensemble average can be readily obtained even in the limit of
isolated resonances. The energy-averaged $S$ matrix is more simply
obtained using a Lorentzian average of width $I$ and given by
\begin{equation}
   \langle S(E) \rangle = S(E+iI) \ ,
   \label{eq:Saverage}
\end{equation}
and $S(E+iI)$ is obtained by replacing $K(E)$ in
Eq.~(\ref{eq:Kmatrix}) or $R(E)$ in Eq.~(\ref{eq:Rmatrix}) by
$K(E+iI)$ or $R(E + i I)$, respectively. That yields the transmission
coefficients in Eq.~(\ref{eq:transmission}).

Monte Carlo calculations based on this approach~\cite{Frohner00} have
been used to define heuristic parametrizations of the width
fluctuation correction factor $W_{a b}$. The results of Hofmann,
Richert, Tepel, and Weidenm\"{u}ller~\cite{Hofmann75,Hofmann80} (HRTW)
are based on the $K$ matrix, those of Moldauer~\cite{Moldauer80} on
the $R$ matrix. The resulting fit formulas for the WFC factor are
collected in the Appendix.

\section{Monte-Carlo simulations}
\label{sec:simulation}

  \subsection{$R$ matrix and $S$ matrix}
  \label{subsec:simRmatrix}

\begin{figure}
  \begin{center} \resizebox{\columnwidth}{!}{\includegraphics{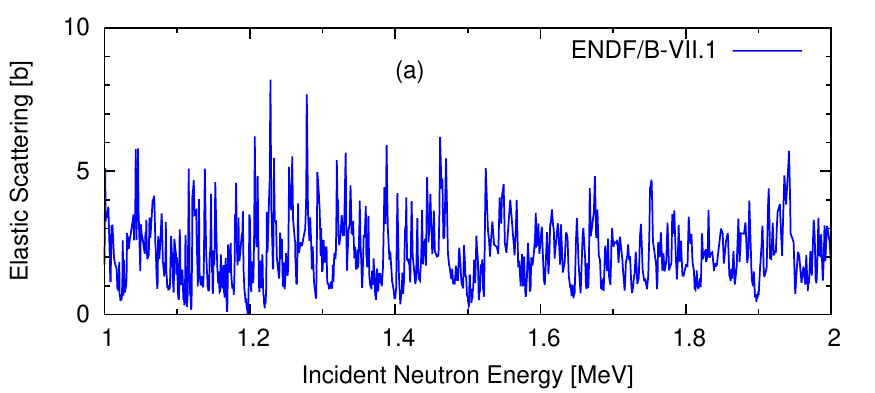}}\\ \resizebox{\columnwidth}{!}{\includegraphics{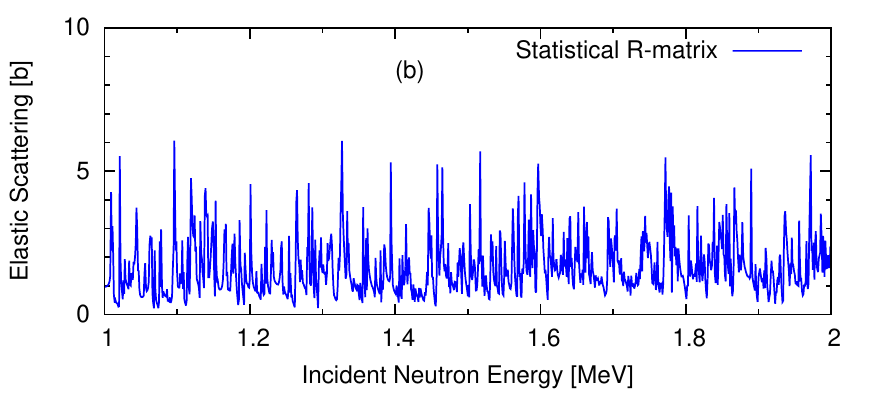}} \end{center}
\caption{(Color
  Online) An example of an elastic scattering cross section generated
  with the statistical $R$-matrix method (bottom panel), compared with
  the real cross section in ENDF/B-VII.1 (top panel), for a
  neutron-induced reaction on $^{56}$Fe in the 1--2~MeV energy
  range.}  \label{fig:Fe56elastic}
\end{figure}

In the 1960s and 70s, the statistical $R$-matrix approach used by
Moldauer~\cite{Moldauer75a, Moldauer75b} offered the only possibility
to use random-matrix ideas in CN scattering. As an example for that
method we show in Fig.~\ref{fig:Fe56elastic} the result of a new
Monte-Carlo simulation of the elastic cross section for neutron
scattering on $^{56}$Fe (bottom panel). This is compared with the real
cross section (upper panel) given in ENDF/B-VII.1 \cite{ENDF7}.  In
the simulation we put $\langle \gamma_c^2 \rangle = 10$~keV for the
$s$ wave, and 200~eV for the higher partial waves ($p$, $d$, and
$f$ waves). The radiative capture channel was ignored. Although the
statistical $R$-matrix calculation cannot reproduce the detailed
structure, the comparison provides information on average properties
and, thus, a useful link between the fluctuating cross section and the
optical model calculations~\cite{Kawano97}.

However, the approach involves a number of parameters, such as the
partial widths, the level density~\cite{Gilbert65, Kawano06}, the
energy range of interest, and so forth. Strict implementation of the
more abstract GOE approach actually removes the need to define these
parameters. This is most easily demonstrated for the case of the $K$
matrix. Upon scaling the energies by the mean level spacing $d$ so
that $E / d \to \epsilon$, $E_\sigma / d \to \epsilon_\sigma$, the
quantities $\epsilon$ and $\epsilon_\sigma$ are dimensionless, and the
spacing distribution of the $\epsilon_\sigma$ is given in terms of the
universal dimensionless correlation functions of the
GOE~\cite{Mehta04}. The expression for $P_W(s)$ in
Eq.~(\ref{eq:wigner}) is an example. Applying the analogous scaling to
the partial width amplitudes, $\gamma_{a \sigma} / d^{1/2} \to
\tilde{\gamma}_{a \sigma}$ generates dimensionless uncorrelated
Gaussian-distributed random variables $\tilde{\gamma}_{a \sigma}$. The
second moments of these quantities are determined by the average
$S$-matrix elements of the optical model. In other words, the scaling
$E / d \to \epsilon$, $E_\sigma / d \to \epsilon_\sigma$, $\gamma_{a \sigma}
/ d^{1/2} \to \tilde{\gamma}_{a \sigma}$ maps the CN scattering problem
onto a GOE scattering problem where the only input parameters are the
number of open channels and the elements $\langle S_{a a} \rangle$ of
the average scattering matrix in each channel. That scattering problem
describes universal chaotic scattering.

\begin{figure*}
  \begin{center}
  \begin{tabular}{cc}
  \resizebox{0.48\textwidth}{!}{\includegraphics{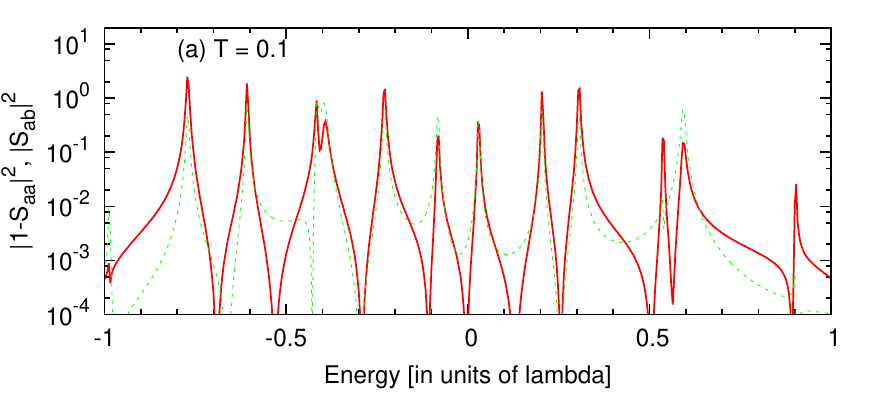}} &
  \resizebox{0.48\textwidth}{!}{\includegraphics{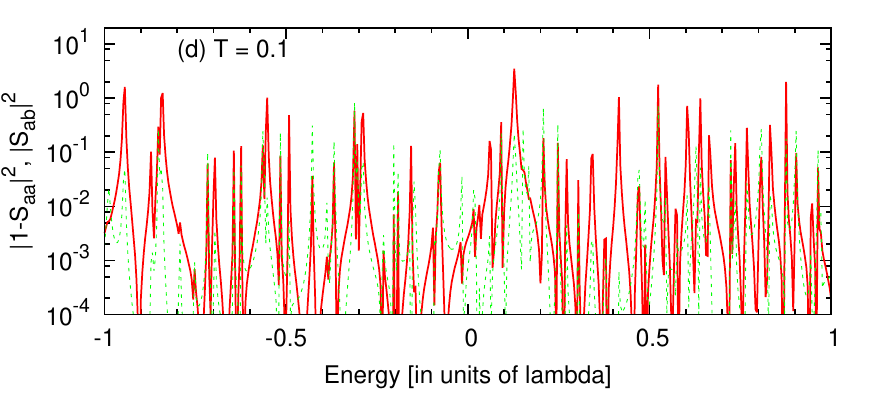}} \\
  \resizebox{0.48\textwidth}{!}{\includegraphics{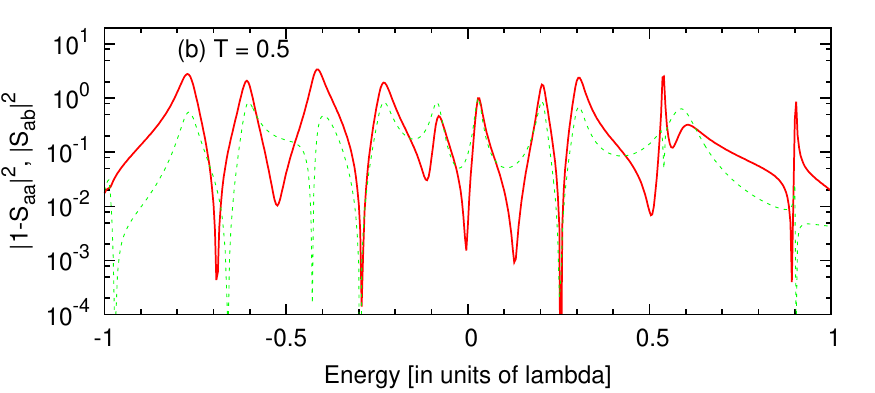}} &
  \resizebox{0.48\textwidth}{!}{\includegraphics{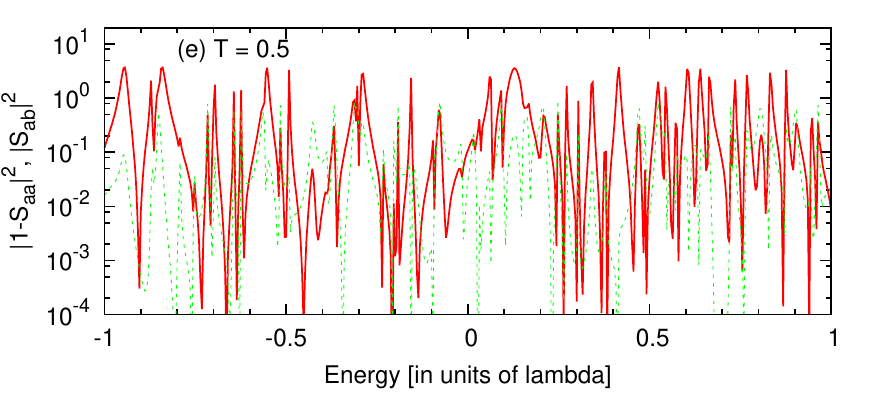}} \\
  \resizebox{0.48\textwidth}{!}{\includegraphics{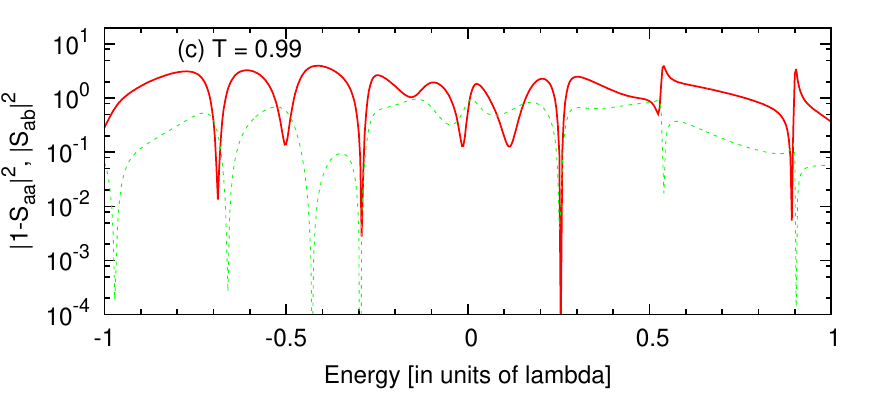}} &
  \resizebox{0.48\textwidth}{!}{\includegraphics{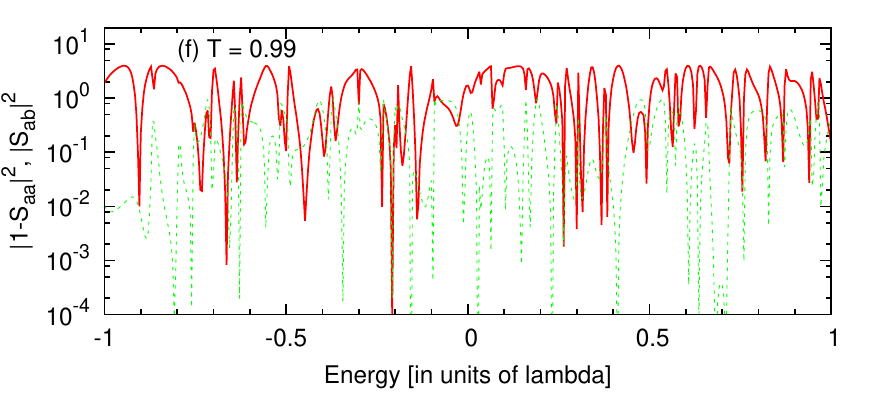}} \\
  \end{tabular}
  \end{center}
  \caption{(Color Online) Generated elastic $|1 - S_{aa}|^2$ and inelastic $|S_{ab}|^2$
    scattering cross sections with the GOE $S$-matrix for three different 
    transmission coefficients $T_a$ of 0.1, 0.5, and 0.99.
    The solid curves are the elastic, and the dotted curves are the inelastic
    cross sections. The left column is for $N=20$ and the right is for the $N=100$ case.}
  \label{fig:crx}
\end{figure*}

For the $S$-matrix simulation, Eqs.~(\ref{eq:Sreal}) and (\ref{prop})
with $H_{\mu \nu}$ replaced by $H^{({\rm GOE})}_{\mu \nu}$ as
\begin{eqnarray}
  S_{a b}^{({\rm GOE})} &=& \delta_{a b}
       - 2 i \pi \sum_{\mu \nu}
         W_{a \mu} \left( D^{-1} \right)_{\mu\nu} W_{\nu b} \ ,
  \label{eq:SGOE}\\
  D_{\mu \nu} &=& E \delta_{\mu \nu}
          - H_{\mu \nu}^{({\rm GOE})} + i \pi \sum_{c} W_{\mu c} W_{c \nu } \ ,
  \label{eq:DGOE}
\end{eqnarray}
serve as starting point. The matrix elements $W_{\mu a}$ obey
Eq.~(\ref{eq:mel}).  The simulation generates an ensemble of $S$
matrices by generating a single set of matrix elements $\{ W_{\mu
a}\}$ combined with a number of realizations of $H^{({\rm GOE})}$. By
construction, all $S$ matrices in the ensemble have for $N \to \infty$
the same mean values. The matrix elements $W_{\mu a}$ are determined
as follows.
Given the elements $G_{\mu a}$ of a coupling strength
matrix $G$ of dimension $N \times \Lambda$, where $\Lambda$ is the
number of channels, diagonalization of the real symmetric matrix
$G^TG$ in channel space with an orthogonal matrix ${\cal O}$ yields
\begin{eqnarray}
  {\cal O}^{-1} G^T G{\cal O} &=& \mbox{diag}(v_a^2) \ ,
  \label{eq:OGO}\\
  W &=& G{\cal O} \ .
  \label{eq:WGO}
\end{eqnarray}
This procedure guarantees that Eqs.~(\ref{eq:mel}) are satisfied. The
eigenvalues $v_a^2$ and $d = \pi \lambda / N$ define $x_a = \pi^2
v^2_a / d$ and these, in turn, the transmission coefficients $T_a$ via
Eq.~(\ref{eq:trans}). Figure~\ref{fig:crx} shows examples of
calculated elastic scattering cross sections for $\Lambda = 2$, $N =
20, 100$, and for three different transmission coefficients $T_a =
0.1$, $0.5$, and $0.99$. To show how the cross section evolves as the
transmission coefficient increases, we fixed the random number sequence so
that the eigenvalues of $H^{\rm (GOE)}$ are the same for the three $T_a$
cases.

  \subsection{Ensemble average}
  \label{subsec:simaverage}

The ensemble average of Eqs.~(\ref{eq:SGOE}) and (\ref{eq:DGOE}) can
be evaluated numerically either by employing the MC technique where
the elements $H^{({\rm GOE})}$ are drawn from a Gaussian distribution,
or by calculating the three-fold integral of Verbaarschot,
Weidenm\"{u}ller, and Zirnbauer \cite{Verbaarschot85}
\begin{widetext}
\begin{eqnarray}
 \overline{S_{ab}^{\rm fl}(E_1) S_{cd}^{{\rm fl}*}(E_2)} &=&
    \frac{1}{8}
    \int_0^\infty d\lambda_1 \int_0^\infty d\lambda_2  \int_0^1 d\lambda  \quad
    \mu(\lambda,\lambda_1,\lambda_2) \nonumber\\
 &\times& e^{-ir(\lambda_1+\lambda_2+2\lambda)}
          {\displaystyle \prod_c}
          \frac{1 - T_c\lambda }{\sqrt{(1+T_c\lambda_1)(1+T_c\lambda_2)}}
          J(\lambda,\lambda_1,\lambda_2) \ ,
   \label{eq:GOE3int}
\end{eqnarray}
where
\begin{eqnarray}
\mu(\lambda,\lambda_1,\lambda_2) 
  &=&  \frac{\lambda(1-\lambda) | \lambda_1 - \lambda_2 |}
       {\sqrt{\lambda_1(1+\lambda_1)} \sqrt{\lambda_2(1+\lambda_2)}
        (\lambda+\lambda_1)^2 (\lambda+\lambda_2)^2 } \ ,  \\
  J(\lambda,\lambda_1,\lambda_2)
  &=& \delta_{ab} \delta_{cd}
      \overline{S}_{aa} \overline{S}_{cc}^* T_a T_c
      \left(
        \frac{ \lambda_1}{1+T_a\lambda_1}
      + \frac{ \lambda_2}{1+T_a\lambda_2}
      + \frac{2\lambda  }{1-T_a\lambda  }
      \right)
      \left(
        \frac{ \lambda_1}{1+T_c\lambda_1}
      + \frac{ \lambda_2}{1+T_c\lambda_2}
      + \frac{2\lambda  }{1-T_c\lambda  }
      \right) \nonumber \\
  &+& (\delta_{ac}\delta_{bd} + \delta_{ad}\delta_{bc}) T_a T_b
      \left\{
         \frac{ \lambda_1(1+\lambda_1)}{ (1+T_a\lambda_1) (1+T_b\lambda_1) }
       + \frac{ \lambda_2(1+\lambda_2)}{ (1+T_a\lambda_2) (1+T_b\lambda_2) }
       + \frac{2\lambda  (1+\lambda  )}{ (1-T_a\lambda  ) (1-T_b\lambda  ) }
     \right\} \ ,
     \label{eq:GOEJ} \\
  r &=& \frac{\pi}{d}(E_2 - E_1) \ .
\end{eqnarray}
\end{widetext}
The triple-integral in Eq.~(\ref{eq:GOE3int}) can be evaluated
numerically by introducing new integration variables
\cite{Verbaarschot86} that avoid singularities in the integrand, and
by the Gauss-Legendre quadrature with the order high enough to obtain
convergence \cite{Hilaire03}. In practical applications we need
$\overline{S_{ab} S_{ab}^*}$ only, so that Eq.~(\ref{eq:GOE3int}) can
be reduced to a slightly simpler form \cite{Hilaire03} as
$\overline{S}_{aa} \overline{S}_{aa}^* = 1 - T_a$. This is not the
case if we have off-diagonal elements in $\langle S\rangle$, or
different energy arguments so that $r \neq 0$.

A benefit of MC is that we are able to explore a larger parameter
space, while Eq.~(\ref{eq:GOE3int}) holds in the limit
$N \rightarrow\infty$. In Eq.~(\ref{eq:GOE3int}) we have replaced the
energy average $\langle|\delta_{ab} - S_{ab}^{({\rm GOE})}|^2\rangle$
by the ensemble average $\overline{|\delta_{ab} - S_{ab}^{({\rm
GOE})}|^2}$. The difference between the two averages is discussed
later. Hereafter we always calculate the ensemble
average unless stated explicitly otherwise.
The average is evaluated at the center of the GOE
eigenvalue distribution, $E=0$. As shown in Fig.~\ref{fig:crx}, the
calculated cross section near $E=0$ for a single realization of
$H^{({\rm GOE})}$ displays chaotic fluctuations. The number of MC
realizations needed to obtain a meaningful average varies from 10,000
to a million, depending on convergence. The criterion used was that
the deviation of the average $S$-matrix from its input value was
sufficiently small, $|\Delta \overline{S}_{ab}| < 10^{-5}$.

\begin{figure}
  \begin{center} \resizebox{\columnwidth}{!}{\includegraphics{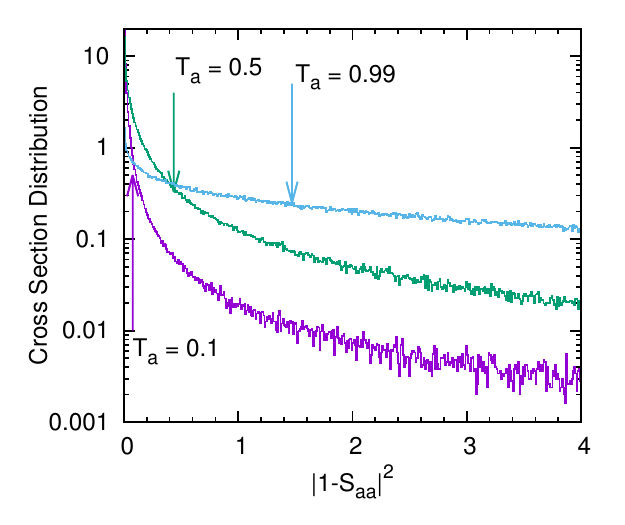}}
  \end{center}
\caption{(Color
  Online) Distribution of elastic scattering cross section at $E=0$ for
  many GOE $S$-matrix realizations. Three cases, $T_a = T_b = 0.1$,
  0.5, and 0.99 are shown. The arrows show
  the actual average values for each distribution.}  \label{fig:sigdist}
\end{figure}

Figure~\ref{fig:sigdist} shows the probability distribution of the
elastic scattering cross section at $E=0$ for $N=100$, $\Lambda=2$,
and three different $T_a=T_b$ values of 0.1, 0.5, and 0.99. The MC
ensemble average values are indicated by the location of the arrows,
e.g., in the case of $T_a=0.99$, the average is 1.47. To compare these
averages with predictions of the statistical model, we have to
subtract the direct part $(1-\Re \overline{S}_{aa})^2$ from the
elastic channel. That gives the average fluctuating part
$\overline{|S^{\rm fl}|^2}$ of 0.660. The Hauser-Feshbach cross
section is
\begin{equation}
  \langle |S_{aa}^{\rm fl}|^2 \rangle = \frac{T_a^2}{T_a + T_b} = 0.495 \ ,
\end{equation}
giving in that case a 25\% smaller elastic cross section. When the GOE
triple-integral of Eq.~(\ref{eq:GOE3int}) is performed for the given
transmission coefficients, the simulated cross sections are recovered.
In Table~\ref{tbl:sigave} we compare the MC results with other
statistical models --- KKM \cite{Kawai73}, HRTW
\cite{Hofmann75,Hofmann80}, Moldauer \cite{Moldauer80}, Ernebjerg and
Herman \cite{Ernebjerg04}, and Kawano and Talou \cite{Kawano13}. In
general, all the statistical models predict the average reasonably
well when $\langle\Gamma\rangle/d$ is large. More comparisons of the
MC generated cross sections with these statistical models can be found in
Ref.~\cite{Kawano13}.

One may argue that the agreement between the GOE triple-integral and
the MC simulation is obvious because the triple-integral is an
analytical form of the ensemble average for Eqs.~(\ref{eq:SGOE}) and
(\ref{eq:DGOE}) in the limit of $N \rightarrow \infty$. We have,
therefore, studied the $N$-dependence of the calculated averages.
Starting with $N=100$, we reduce the number of resonances and compare
the ensemble average of the elastic cross section with the
triple-integral results. Surprisingly the triple-integral still gives
very accurate average values even if $N=3$. Averaging over a few
resonances is certainly an extreme case, and is not realistic.

\begin{table*}
 \caption{Comparison of numerical average $\overline{|S_{aa}^{\rm fl}|^2}$
    for some cases of $T_a= T_b=0.1$, 0.5, and 0.99,
    with the statistical models --- 
    Hauser-Feshbach \cite{Hauser52}, KKM \cite{Kawai73}, 
    HRTW \cite{Hofmann80}, Moldauer \cite{Moldauer80}, GOE \cite{Verbaarschot85},
    Ernebjerg-Herman \cite{Ernebjerg04},
    and Kawano-Talou \cite{Kawano13}.}
 \label{tbl:sigave}
\begin{tabular}{crrrrrr}
\hline
$T_a$ & \multicolumn{2}{c}{0.1} & \multicolumn{2}{c}{0.5} & \multicolumn{2}{c}{0.99} \\
\hline
                & Elastic  & Inelastic& Elastic & Inelastic& Elastic& Inelastic\\
\hline
MC simulation   & 0.0733  &  0.0261  &  0.351  &  0.149  &  0.660 &  0.330\\
\hline
Hauser-Feshbach & 0.0500  &  0.0500  &  0.250  &  0.250  &  0.495 &  0.495\\
KKM             & 0.0662  &  0.0332  &  0.333  &  0.167  &  0.660 &  0.330\\
HRTW            & 0.0737  &  0.0257  &  0.352  &  0.147  &  0.661 &  0.330\\
Moldauer        & 0.0734  &  0.0260  &  0.349  &  0.150  &  0.665 &  0.325\\
GOE             & 0.0734  &  0.0260  &  0.351  &  0.148  &  0.661 &  0.330\\
Ernebjerg-Herman & 0.0742  &  0.0252  &  0.366  &  0.134  &  0.681 &  0.310\\
Kawano-Talou     & 0.0735  &  0.0259  &  0.351  &  0.148  &  0.661 &  0.330\\
\hline
\end{tabular}
\end{table*}

\section{Validation of statistical models}
\label{sec:statistical}

  \subsection{Energy average versus ensemble average}
  \label{subsec:stataverage}

There are three ways to calculate averages: (a) the ensemble average
can be performed analytically in the limit $N \to \infty$, which is
given in Eq.~(\ref{eq:GOE3int}), (b) the ensemble average can be
performed numerically using the MC simulations for finite $N$, and (c)
the average is taken over energy and calculated for a single
realization of the ensemble. Method (c) is the only way to perform
averages over actual data. Such averages define the optical
model.  Obviously it is highly important to know whether (and if so,
when) these averages agree.

Let $w(E_0, E, I)$ be the weight function centered at energy
$E_0$ with width $I$ used to define the average over energy $E$. In
what follows $w(E_0,E, I)$ is taken to be a Lorentzian. Our aim is to
know under which circumstances the equality
\begin{equation}
   \int_{-\infty}^{+\infty} w(E_0, E, I) S(E) dE = \overline{S}(E_0)
\end{equation}
holds. Since there is no analytical way to investigate that relation,
we ask when the weaker condition
\begin{equation}
  \overline{ \left| \langle S \rangle - \overline{S} \right|^2 } = 0
\label{condition}
\end{equation}
is fulfilled \cite{Brody81}. It is straightforward to show that
Eq.~(\ref{condition}) is equivalent to
\begin{eqnarray}
  &~& \int_{-\infty}^{+\infty} dE_1 w(E_0, E_1, I)
      \int_{-\infty}^{+\infty} dE_2 w(E_0, E_2, I) \nonumber \\
  &~& \overline{S^{\rm fl}(E_1) S^{{\rm fl} *}(E_2)} = 0 \ ,
\end{eqnarray}
where the two-point function $\overline{S^{\rm fl}(E_1) S^{{\rm fl}*}(E_2)}$
is given by Eq.~(\ref{eq:GOE3int}).

The average two-point function $\overline{S^{\rm fl}(E_1) S^{{\rm
fl}*}(E_2)}$ involves two $S$ matrices at energies $E_1$ and $E_2$.
Because of the very weak energy dependence of the average $S$ matrix
we approximate $\overline{S}(E_1) \simeq \overline{S}(E_2)$ and
evaluate both at $E=0$. Then the energies $E_1$ and $E_2$ in the
two-point function appear only in the oscillating term
\begin{equation}
  \exp\left\{-ir(\lambda_1+\lambda_2+2\lambda)\right\} \ , \quad
  r = \frac{\pi}{d}(E_2-E_1) \ .
  \label{eq:exp}
\end{equation}
We assume that the level spacing $d = \pi\lambda/N$ is
independent of energy. We limit ourselves to the case where all
transmission coefficients are equal and given by $T_a$. We perform the
energy averages using Lorentzians centered at zero,
\begin{equation}
  w(E,0,I) = \frac{I}{\pi} \frac{1}{E^2 + I^2}
  \label{eq:Lorentzian}
\end{equation}
with width $I$ specified in units of $d / \pi$. We define
\begin{eqnarray}
  L(T_a,\Lambda,I) &=&
      \int_{-\infty}^\infty \int_{-\infty}^\infty
      w(E_1,0,I) w(E_2,0,I) \nonumber \\
      &\times& R(E_2-E_1;T_a,\Lambda) \ dE_1 dE_2 \ ,
  \label{eq:Lfunc}
\end{eqnarray}
where
\begin{equation}
  R(E_2-E_1;T_a,\Lambda)  =
   \frac{ \Re\left\{\displaystyle\overline{S^{\rm fl} S^{{\rm fl}*}}(|E_2 - E_1|) \right\}}
        {\displaystyle\overline{|S|^2}(0)} \ .
\end{equation}
We use the real part only because integration over the imaginary part in 
Eq.~(\ref{eq:Lfunc}) yields zero.

\begin{figure}
 \resizebox{\columnwidth}{!}{\includegraphics{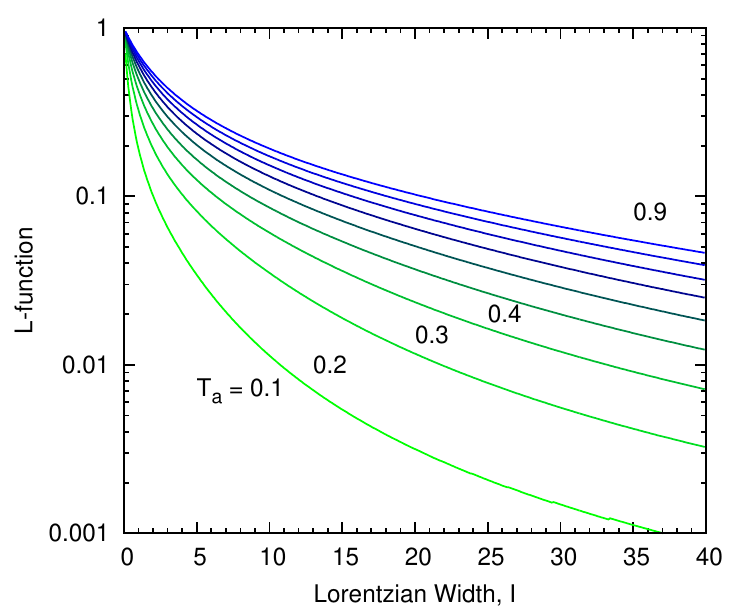}}
 \caption{(Color Online) $L(T_a,\Lambda,I)$ as defined in
   Eq.~(\ref{eq:Lfunc}) versus the Lorentzian width $I$ for $\Lambda
   = 10$ channels and for different values of the transmission
   coefficient $T_a$. From the lowest to the highest curve $T_a$
   changes from 0.1 to 0.9 in steps of width 0.1.}
 \label{fig:corint_m10}
\end{figure}

\begin{figure}
\resizebox{\columnwidth}{!}{\includegraphics{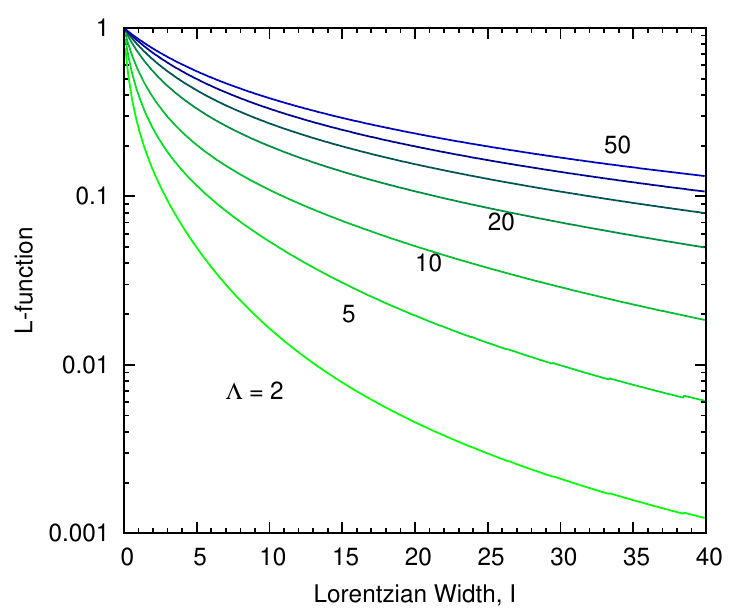}}
\caption{(Color Online) $L(0.5,\Lambda,I)$ as defined in
  Eq.~(\ref{eq:Lfunc}) versus the the Lorentzian width $I$ for $T_a =
  0.5$ and for different channel numbers $\Lambda$. From the lowest to
  the highest curve the values of $\Lambda$ are 2, 5, 10, 20, 30, 40, and
  50.}
\label{fig:corint_t05}
\end{figure}

\begin{figure}
\resizebox{\columnwidth}{!}{\includegraphics{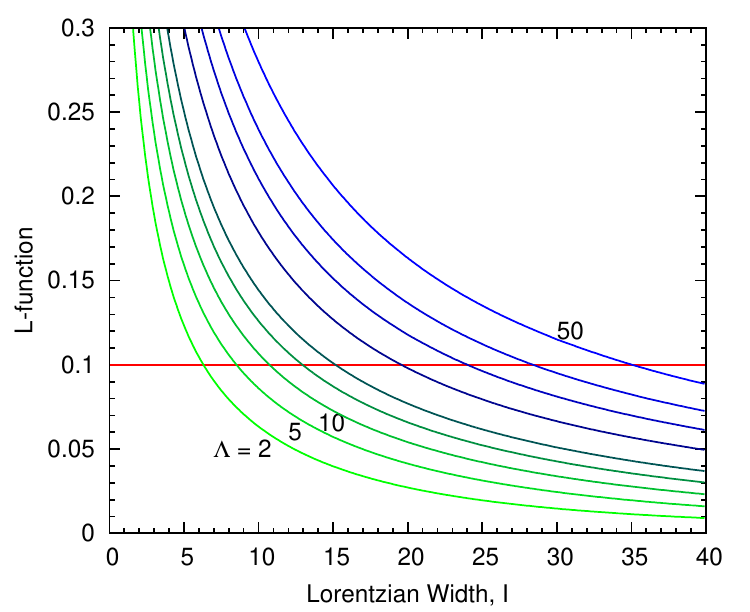}}
\caption{(Color Online) $L(0.99,\Lambda,I)$ as defined in
  Eq.~(\ref{eq:Lfunc}) versus the Lorentzian width $I$ for $T_a = 0.99$ and
  for different channel numbers $\Lambda$. From the lowest to the highest 
  curve the values are $\Lambda$ = 2, 3, 4, 5, 6, 8, 10, 12, and 15. The
  horizontal line shows $L=0.1$.}
 \label{fig:corint_t99}
\end{figure}

\begin{figure}
\resizebox{\columnwidth}{!}{\includegraphics{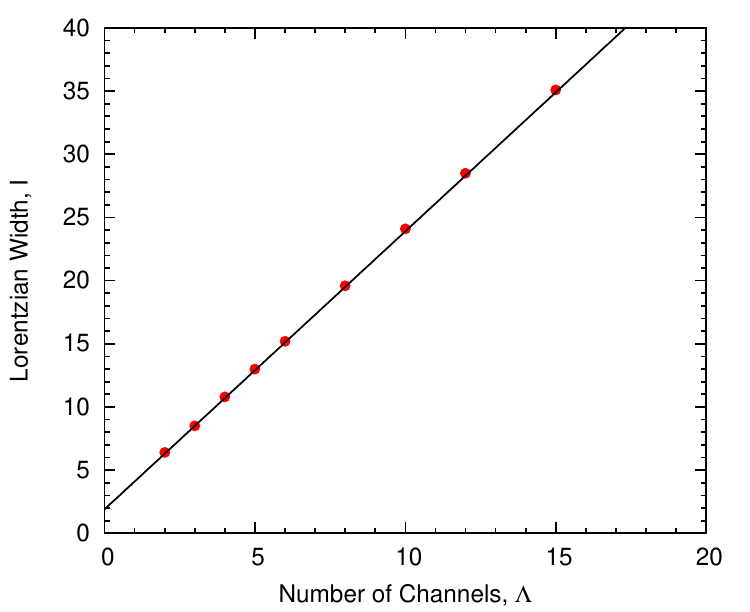}}
\caption{(Color Online) The symbols show for which values of the
  Lorentzian width $I$ and channel number $\Lambda$ the function
  $L(0.99,\Lambda,I)$ attains the value $0.1$. The line is a
  least-squares fit to the symbols; $I(L=0.1) \simeq 2.2\Lambda + 1.9.$}
  \label{fig:width01}
\end{figure}

Our results for the elastic channel are displayed in
Figs.~\ref{fig:corint_m10} to \ref{fig:width01}.
Figure~\ref{fig:corint_m10} shows the function $L$ of
Eq.~(\ref{eq:Lfunc}) versus $I$ for $\Lambda = 10$ and for values of
$T_a$ ranging from 0.1 to 0.9. As expected, $L$ decreases as $I$
increases so that ensemble average and energy average agree when the
Lorentzian width $I$ is sufficiently large. To get the same accuracy
larger values of $T_a$ require larger widths $I$. The dependence of
$L$ on channel number $\Lambda$ is shown versus $I$ in
Fig.~\ref{fig:corint_t05} for $T_a=0.5$ (logarithmic scale) and in
Fig.~\ref{fig:corint_m10} for $T_a = 0.99$ (linear scale). Larger
values of $\Lambda$ require larger values of $I$, the slowest decrease
occurring for the strong-absorption case where $T_a$ is close to
unity. For the strong-absorption case $T_a=0.99$,
Fig.~\ref{fig:width01} shows the values of $I$ versus channel
number for which $L = 0.1$. The result is a clear linear dependence
\begin{equation}
  I(L = 0.1) \simeq 2.2\Lambda + 1.9 \ .
\end{equation}

In the Ericson regime $\sum_a T_a \gg 1$ or, for equal
transmission coefficients in all channels, $T_a \Lambda \gg 1$, the
autocorrelation function is known analytically. The real part is a
Lorentzian with denominator $r^2 + \Gamma^2$ where the total width is
given by $\Gamma = (d / 2 \pi) \sum_a T_a$. For large $I$ the function
$L$ falls off with $(2 I)^{- 1}$. We have $L = 0.1$ for $I \approx
5 \Gamma$.

 The rate of decrease of $L$ versus $I$ depends on $T_a$ and
$\Lambda$.  Using our results we can nevertheless draw some general
conclusions concerning neutron-induced reactions at low energy. In the
domain of isolated resonances the number of channels is effectively
small ($\gamma$ channels are numerous but extremely weak
individually).  Here the $L$-function becomes $\sim$ 0.1 or less when
$I$ is larger than 10 or so. A value of $I=100$ corresponds to $100 d
/\pi \sim 30d$. Hence the $L$-function will be sufficiently small when
the energy-averaging interval is one or two orders of magnitude larger
than the average resonance spacing $d$. In the Ericson regime that
same statement applies with $d$ replaced by $\Gamma$, the average
total resonance width.

  \subsection{Asymptotic value at strong-absorption limit}
  \label{subsec:statstrong}

\subsubsection{Elastic enhancement factor in Ericson limit}

In the strong-absorption or Ericson limit $\sum_a T_a \gg 1$,
Eq.~(\ref{eq:GOE3int}) yields $W_a = 2$ for the elastic
enhancement factor or, equivalently, $\nu_a = 2$ for the channel
degree-of-freedom \cite{Verbaarschot86}. Explicitly we have
\begin{equation}
  \langle \sigma_{a b} \rangle
  = \frac{(1 + \delta_{a b}) T_a T_b}{\sum_c T_c} + \ldots \ .
\label{eq:HFexpansion}
\end{equation}
The dots indicate terms of order $(\sum_c T_c)^{- 2}$ or higher. The
term of leading order is the Hauser-Feshbach result with an elastic
enhancement factor of two. Most statistical models agree with that
result. An exception is the model by Moldauer, which has an asymptotic
value of $\nu_a = 1.78$. Although Moldauer's heuristic method to
obtain Eq~(\ref{eqA:MoldauerSyst}) in the Appendix is somewhat similar
to the MC technique we adopt here, there is a notable difference
between the two approaches. In the MC approach we perform the ensemble
average over the elements of the Hamiltonian $H_{\mu \nu}^{({\rm
GOE})}$. Moldauer's statistical $R$-matrix model has two independent
inputs: the decay widths drawn from the Porter-Thomas distribution,
and the level spacing sampled from the Wigner distribution in
Eq.~(\ref{eq:wigner}).

\subsubsection{Decay amplitude distribution}

Our aim is to reproduce Moldauer's lower asymptotic value by modifying
the MC sampling method. Before doing that, we show the distribution of
the width amplitudes $\sqrt{\pi} \gamma_{a \sigma}$ when we rewrite our
stochastic $S$-matrix of Eq.~(\ref{eq:SGOE}) in an equivalent
form \cite{Verbaarschot85,Mitchel10}
\begin{eqnarray}
  K_{ab}(E) &=& \sum_\sigma \frac{\tilde{W}_{a\sigma} \tilde{W}_{\sigma b}}{E-E_\sigma} \ ,
  \label{eq:KGOE} \\
  \tilde{W}_{\sigma a} &=& \sqrt{\pi} \sum_\nu {\cal O}_{\sigma\nu} W_{\nu a} \ ,
  \label{eq:Kwidth} \\
  {\cal O}^{-1} H^{(\rm GOE)} {\cal O}
   &=& \mbox{diag}(E_\sigma) \ ,
  \label{eq:GOEdiag}
\end{eqnarray}
where $E_\sigma$ is the eigenvalue of $H^{(\rm GOE)}$. In this form the
width amplitudes $\tilde{W}_{a \sigma} = \sqrt{\pi} \gamma_{a \sigma}$
are uncorrelated Gaussian-distributed random variables with zero mean
values and the standard deviation. We produced the distributions of
$\tilde{W}$ for the case $N=100$, $\Lambda=2$, and three values of
$T_a=0.1$, 0.5, and 0.99.

\begin{figure}
 \begin{center}
   \resizebox{\columnwidth}{!}{\includegraphics{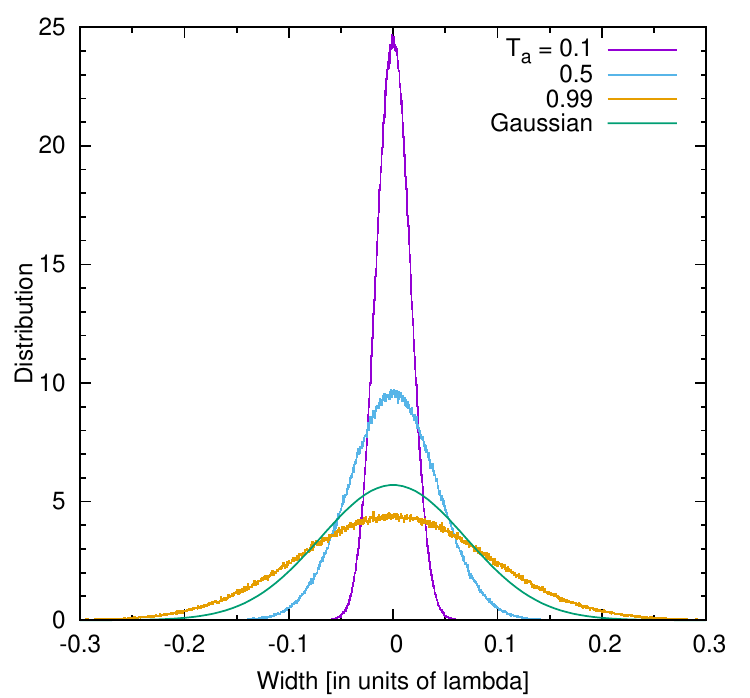}}
 \end{center}
\caption{(Color Online) Distribution of decay amplitudes
  $\gamma_{a \sigma}$, when the GOE $S$ matrix is written in the $K$-matrix
  form. The histograms correspond to $T_a=0.1$, 0.5, and 0.99, respectively.
  For $T_a=0.99$, we compare the Gaussian distribution with the one obtained
  from the standard deviation in Eq.~(\ref{eq:gaussstd2}).}
\label{fig:widdist}
\end{figure}

\begin{figure}
 \begin{center}
   \resizebox{\columnwidth}{!}{\includegraphics{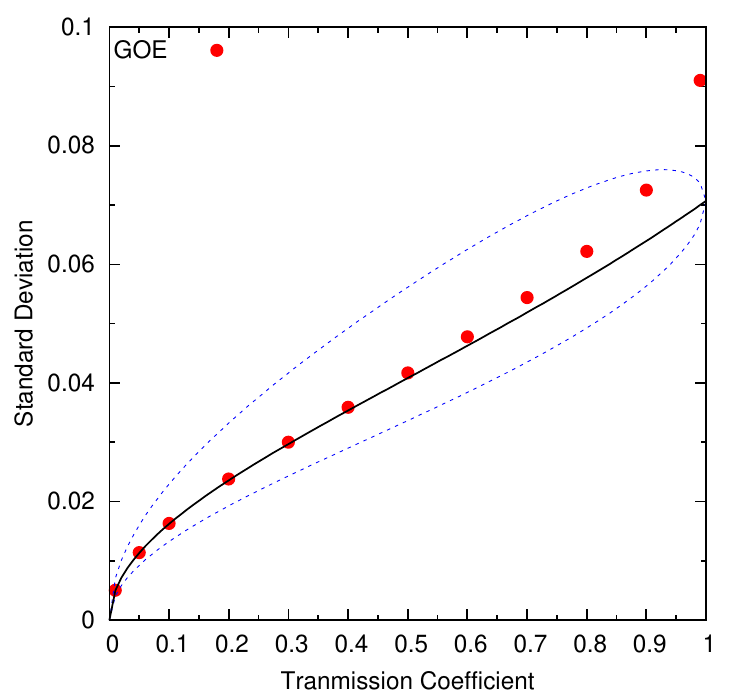}}
 \end{center}
\caption{(Color Online)
  Standard deviation of GOE decay amplitude distribution
  as a function of transmission coefficient. The dashed curves
  are Eq.~(\ref{eq:moment}) and the solid curve is Eq.~(\ref{eq:gaussstd2}).}
\label{fig:widdist_sigma}
\end{figure}

The width distributions are shown in Fig.~\ref{fig:widdist} for the
elastic channel. Because we used the same transmission for both
channels, the distribution for the elastic and inelastic channels are
identical. Figure~\ref{fig:widdist_sigma} shows the standard deviation
$\sigma_a$ for each Gaussian for various $T_a$.

The second moment of Gaussian distribution is given by \cite{Verbaarschot85}
\begin{equation}
  \sigma_a^2 = \frac{d}{\pi} \frac{T_a}{2-T_a \pm \sqrt{1-T_a}} \ ,
 \label{eq:moment}
\end{equation}
which is shown by the two dashed curves in Fig.~\ref{fig:widdist_sigma}.
The sign ambiguity in Eq.~(\ref{eq:moment}) is caused by the fact that
there are two values of $\overline{S}_{a a}$ with opposite signs that
yield the same value of $T_a = 1 - |\overline{S}_{a a}|^2$.

\subsubsection{Emulating Moldauer's calculation}

Moldauer's $K$ matrix (see Section \ref{subsec:theoryS}) can be
written as
\begin{equation}
   K_{ab}^{\rm M}(E) = 
   \delta_{ab} \Re K_a^{0}
    + \sum_\sigma \frac{w_{a\sigma} w_{\sigma b}}{E-E_\sigma} \ .
  \label{eq:KMoldauer}
\end{equation}
The elements $K_a^{0}$ of the elastic background matrix and the
variances of the amplitudes $w$ are determined by the energy-averaged
$S$ matrix. For $K_a^{0}$ we have
\begin{equation} 
K^0 = i \frac{1 -  S^{(\rm GOE)}(E+iI)}
             {1 +  S^{(\rm GOE)}(E+iI)} \ ,
 \label{eq:Kaverage}
\end{equation}
showing that $K_a^{0}$ is determined by the transmission coefficient
$T_a$. Since $\Im S(E+iI) \sim 0$, we may omit the background term
$\Re K_a^{0}$. When we view the $K$-matrix as an $R$-matrix, $\Im
K(E+iI)$ is the pole strength $2\pi \overline{\gamma_a^2}/d$,
therefore the second moment for the distribution of the widths
$w_{a \sigma}$ reads
\begin{equation}
  \sigma_a^2 = 2\pi\overline{\gamma_a^2}
             = \frac{d}{\pi} \left|\Im K_a^{0}\right| \ .
  \label{eq:gaussstd1}
\end{equation}
The elastic enhancement factor $W_a$ can be defined only when all
channels are identical, $T_a = T_b = \ldots = T_\Lambda$. That is the
case we address.

The calculation of the ensemble average of Eq.~(\ref{eq:KMoldauer})
proceeds as follows. First we generate $S$ of Eq.~(\ref{eq:SGOE}), and
convert it into $K$ via Eq.~(\ref{eq:SKmatrix}). As
Moldauer performed in Ref.~\cite{Moldauer80}, we use $K(E+iI)$ and
Eq.~(\ref{eq:gaussstd1}) to determine the average widths of the decay
amplitudes. The latter are then sampled from Gaussians with widths
$\sigma_a$, independently of the GOE eigenvalues. The Lorentzian
average width $I$ is taken to be 0.2 $\lambda$. We extract the elastic
enhancement factors and compare with the standard GOE simulation that
is described in Sec. \ref{subsec:simaverage}.

The elastic enhancement factor $W_a$ is calculated as
\begin{eqnarray}
 W_{aa} &=& \frac{ \overline{| S_{aa}^{{\rm fl}} |^2}}
                { \sigma_{aa}^{{\rm HF}} } \ , \qquad
 \sigma_{aa}^{{\rm HF}} = \frac{T_a^2}{\Lambda T_a} \ ,\\
 W_a &=& \frac{(\Lambda-1)W_{aa}}{\Lambda - W_{aa}} \ ,\qquad
 \nu_a = \frac{2}{W_a - 1} \ .
\end{eqnarray}
We calculate $K(E+iI)$ for each realization of the GOE $S$-matrix.
Therefore, the ensemble average of Eq.~(\ref{eq:KMoldauer}) converges
slowly. In addition, simulations for very large values of $N$ or
$\Lambda$ are not feasible in general. We chose $N=200$, $\Lambda=5$,
10, 20, and 30. The transmission coefficients are 0.25 and 0.75.
These combinations roughly cover Moldauer's numerical study of the
strong-absorption cases.

\begin{figure}
  \begin{center}
    \resizebox{\columnwidth}{!}{\includegraphics{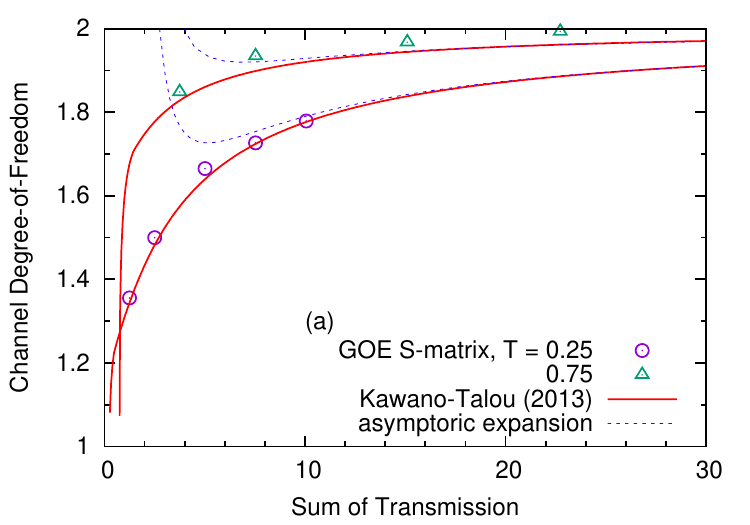}}\\
    \resizebox{\columnwidth}{!}{\includegraphics{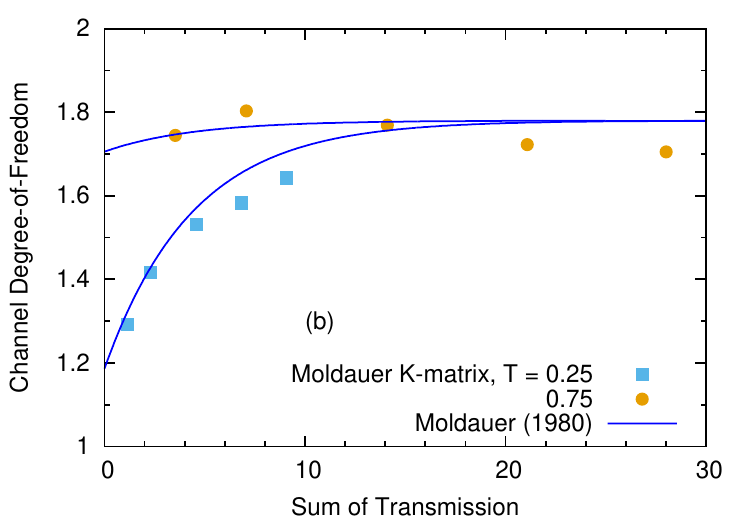}}
  \end{center}
\caption{(Color Online) Channel degree-of-freedom values $\nu_a$ as functions
  of $\sum_a T_a$. The symbols are the Monte-Carlo simulation results, see
  text. The solid curves in the top panel are from Eq.~(\ref{eqA:LANLsyst}),
  and the dotted curves are from Eq.~(\ref{eqA:asympt}) for $T_a=0.25$ and 0.75.
  The curves in the bottom panel show Moldauer's
  systematics given by Eq.~(\ref{eqA:MoldauerSyst}) for the same set of $T_a$.}
  \label{fig:emumold}
\end{figure}

The values of $\nu_a$ versus $\sum_a T_a$ obtained in that way
are compared with the GOE result in Fig.~\ref{fig:emumold}. The
symbols in the upper panel show the results of the standard GOE
simulation, those in the lower panel the results of the simulation
described in the previous paragraph. The curves in the upper panel
represent Eqs.~(\ref{eqA:LANLsyst1}), those in the lower panel
represent Eq.~(\ref{eqA:MoldauerSyst}), both for the cases $T_a =
0.25$ and 0.75. These equations are meant to approximate $\nu_a$ for
given values of the transmission coefficients. The results of the GOE
simulation are well represented by Eq.~(\ref{eqA:LANLsyst1}) which has
the asymptotic value of 2 in the strong-absorption limit. The MC
simulation that uses Eq.~(\ref{eq:gaussstd1}) tends to give lower
$\nu_a$ values, similar to Moldauer's findings.

A plausible explanation of this discrepancy relates to the
determination of the decay amplitude via Eq.~(\ref{eq:gaussstd1}).
Since
\begin{equation}
 \Im K_a^{0} = \frac{T_a}{2 - T_a} \ ,
\end{equation}
the widths in Moldauer's approach have a second moment given by
\begin{equation}
  \sigma_a^2 = \frac{d}{\pi} \frac{T_a}{2-T_a} \ ,
  \label{eq:gaussstd2}
\end{equation}
which is shown in Fig.~\ref{fig:widdist_sigma} by the solid
curve.
Comparison with Eq.~(\ref{eq:moment}) shows that this is correct only
for small values of $T_a$. Discrepancies arise for $T_a \approx 1$. In
Fig~\ref{fig:widdist} we compare for $T_a=0.99$ the distribution of
widths using for the second moment the correct
expression~(\ref{eq:moment}) with the one obtained from Moldauer's
equation~(\ref{eq:gaussstd2}). (We do not show the $T_a=0.1$ and 0.5
cases because they perfectly overlap with the exact values). We note
that Moldauer's approach gives a slightly narrower distribution. We
suspect that this is the root of Moldauer's incorrect asymptotic value
for $\nu_a = 1.78$.

\subsubsection{Asymptotic expansion}

The next-to-leading-order term of Eq.~(\ref{eq:HFexpansion}) is
given by an asymptotic expansion of Eq.~(\ref{eq:GOE3int}) in inverse
powers of $\sum_c T_c$ \cite{Weidenmuller84,Verbaarschot86}, which is
also given in Appendix. This is shown by the dashed curves in
Fig.~\ref{fig:emumold} (a). The asymptotic expansion approximates the
GOE triple-integral very well, when $\sum_cT_c > 10$.  This might be
practically useful in the strong-absorption limit, in particular when
the number of open channels is so large that calculation of the GOE
triple-integral becomes extremely difficult.

  \subsection{Very weak entrance channel}
  \label{subsec:statweak}

\begin{figure}
  \begin{center}
  \resizebox{\columnwidth}{!}{\includegraphics{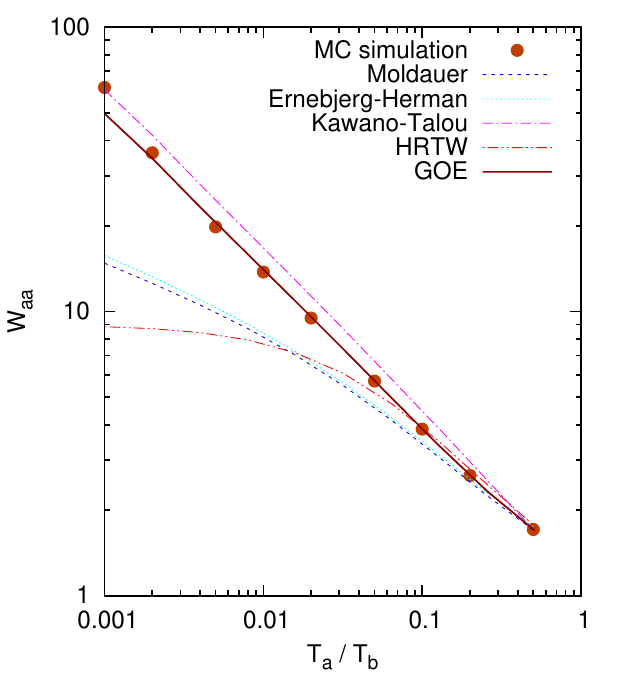}}
  \end{center}
  \caption{(Color Online) Ratio of the elastic cross section to the
    Hauser-Feshbach prediction as a function of the ratio $T_a/T_b$ of
    transmission coefficients. The symbols are the MC simulation
    results. The curves are predictions by various statistical models.}
  \label{fig:smallentrance}
\end{figure}

An extreme case where all the statistical models fail is reported in
Ref.~\cite{Kawano13}. When there are few open channels with either
very small or very large transmission coefficients, none of the width
fluctuation models reproduces the GOE results. That was also discussed
by Moldauer \cite{Moldauer76} as the total width fluctuation, and his
numerical study shows a strong enhancement in the elastic channel. We
performed the GOE simulation for the case of $N=100$, $\Lambda=2$ and
$T_a/T_b \ll 1$. The calculated width fluctuation correction factor
$W_{aa}$, which is the ratio of the elastic channel cross section to
the Hauser-Feshbach cross section, is shown in
Fig.~\ref{fig:smallentrance}. Since the GOE triple-integral is correct
for all values of $\Lambda$ and $T_a$, the MC simulation perfectly
agrees with GOE, except some deviation seen at very small $T_a/T_b$
values, due to numerical instability.

Few-channel cases with very different values of the transmission
coefficients are very special and hard to realize in practice. A
photo-induced reaction that creates a compound nucleus just above
neutron threshold could be a case in point. However, since almost all
incoming flux goes to the neutron channel and to the other gamma
channels, the photon compound elastic cross section is tiny even if it
is enhanced by a factor of 50. That is why it might be difficult to
confirm the strong enhancement in the elastic channel experimentally.

\section{Direct reactions}
\label{sec:direct}

  \subsection{Engelbrecht-Weidenm\"{u}ller transformation}
  \label{subsec:directtheory}

So far it was assumed that the average $S$ matrix is
diagonal. That assumption fails when some channels are strongly
coupled. In practice that happens, for instance, when collective
states in the target nucleus are excited by an incident nucleon (a
direct reaction). In such cases, the average $S$ matrix is not
diagonal. The unitarity of the scattering matrix imposes strong
constraints on the scattering amplitudes. As a consequence, directly
coupled channels cause correlations between the resonance amplitudes
in those channels. That is why the calculation of the average
compound-nucleus cross section in the presence of direct reactions 
has been a long-standing problem.

When $\langle S \rangle$ is not diagonal, the definition of the
transmission coefficients $T$ must be generalized. That is done using
Satchler's transmission matrix \cite{Satchler63}
\begin{equation}
  P_{ab}
  = \delta_{ab} - \sum_c \langle S_{ac}\rangle \langle S_{bc}^* \rangle \ .
  \label{eq:Pmat}
\end{equation}
In the strong-absorption limit, Kawai, Kerman and McVoy (KKM)
\cite{Kawai73} expressed the compound-nucleus cross section in
terms of the matrix $P$ (see Eqs.~(\ref{eqA:Xmatrix}) and
(\ref{eqA:KKM})). Actual calculations using KKM including the direct
channels are, unfortunately, very limited, e.g. Refs.\cite{Arbanas08}
and \cite{Kawano08}.

In practical calculations, an often-used approximate way to
include the direct reaction in the statistical model consists in
redefining the transmission coefficients so as to take account of some
direct reaction contribution,
\begin{equation}
   T'_a = 1 - \sum_c |\langle S_{ac}\rangle \langle S_{ac}\rangle^* |^2 \ .
  \label{eq:Tmod}
\end{equation}
The sum of the modified transmission coefficients $T'_a$ equals ${\rm
Tr}(P)$. Therefore, it is reasonable to expect that GOE
cross-section calculations using the modified transmission
coefficients $T'_a$ as input parameters as done in
Ref.~\cite{Kawano09} may not be far off the mark. In comparison with
the exact approach introduced below, the method greatly simplifies the
calculations. However, a quantitative validation of the
simplification~(\ref{eq:Tmod}) and an understanding of its limitations
are still needed.

The following rigorous treatment of the direct reaction was
proposed by Engelbrecht and Weidenm\"{u}ller (EW) \cite{Engelbrecht73}.
Since $P$ is hermitian, $P$ can be diagonalized by a unitary matrix
\begin{equation}
   (UPU^\dag)_{ab} = \delta_{ab} p_a \ , \qquad 0 \le p_a \le 1 \ .
\end{equation}
The transformation $U$ also diagonalizes the average scattering
matrix,
\begin{equation}
   \langle\tilde{S}\rangle = U \langle S \rangle U^T \ {\rm with} \
   \langle\tilde{S}\rangle_{a b} = \delta_{a b}
   \langle\tilde{S}\rangle_{a a} \ .
\end{equation}
In the diagonal basis of $P$, the transmission coefficients are given by
\begin{equation}
   p_a = 1 - | \langle\tilde{S}_{aa}\rangle |^2 \ .
\end{equation}
In that basis, the decay amplitudes in different channels are
statistically uncorrelated, and the calculation of
$\overline{\tilde{S}_{pq} \tilde{S}_{rs}^*}$ proceeds as described
above for the case without direct reactions, with $p_a$ as input
parameters. The result must be transformed back to the physical
channels. That gives \cite{Hofmann75}
\begin{equation}
  \overline{|S_{ab}|^2}
      = \sum_{pqrs} U_{pa}^* U_{qb}^* U_{ra} U_{sb}
        \overline{\tilde{S}_{pq} \tilde{S}_{rs}^*} \ .
  \label{eq:backtrans}
\end{equation}

Moldauer demonstrated the impact of the EW transformation
numerically \cite{Moldauer75b}. He argued that the flux into the
strongly coupled inelastic channels is enhanced. Capote {\it et
al.} \cite{Capote14} demonstrated that enhancement by applying the
coupled-channels code ECIS \cite{ECIS} to neutron scattering off
$^{238}$U. Although ECIS is capable of performing the EW
transformation, it has some approximations and limited functionality,
particularly for calculating the neutron radiative capture and fission
channels. The EW approach uses only the average $S$ matrix as input
and facilitates showing how direct reactions impact on the compound
nucleus.

A closed form of the average cross section based on the GOE
triple-integral formula that takes the EW transformation into account,
was derived by Nishioka, Weidenm\"{u}ller, and
Yoshida \cite{Nishioka89}. However, the computation might be
impractical. We follow the EW transformation step-by-step from
Eq.~(\ref{eq:Pmat}) to Eq~(\ref{eq:backtrans}). The result allows us
to estimate uncertainties due to the approximation Eq.~(\ref{eq:Tmod}).

  \subsection{Ensemble average using EW transformation}
  \label{subsec:directaverage}

To implement direct reactions, one may use, for instance, the
pole expansion of the $S$ matrix. We find it simpler to employ the
$K$-matrix as in Eq.~(\ref{eq:Kmatrix}). We allow for a direct
background by writing
\begin{equation}
  K_{ab}(E) =
  K^{(0)}_{a b} + \sum_\sigma \frac{\tilde{W}_{a\sigma} \tilde{W}_{\sigma b}}
  {E-E_\sigma}
  \label{eq:KwithBG}
\end{equation}
where the elements of the background matrix $K^{(0)}$ serve as
parameters. When $K$ is real and symmetric, $S$ is automatically
unitary.

We consider a case with direct coupling between two channels only.
The background matrix $K^{(0)}$ is
\begin{equation}
 K^{(0)} =
 \left(
   \begin{array}{cccc}
     k_{aa} & k_{ab} & 0 & \cdots \\
     k_{ab} & k_{bb} & 0 & \cdots \\
     0     & 0     & 0 & \cdots \\
     \vdots & \vdots & \vdots  \\
   \end{array}
 \right)  \ .
 \label{eq:Kbg}
\end{equation}
For the sake of simplicity, we take $k_{aa} = k_{ab} = k_{bb} = k_0$,
where $k_0$ is real. The average $S$ matrix is
\begin{equation}
  \overline{S}
  = \frac{1 - i K^{(0)} + \pi \langle \tilde{W}_a \tilde{W}_b \rangle}
         {1 + i K^{(0)} - \pi \langle \tilde{W}_a \tilde{W}_b \rangle} \ .
\end{equation}
The amplitudes $\tilde{W}_{a \sigma}$ are zero-centered
Gaussian-distributed random variables, uncorrelated for $a \neq b$.
The parameters then are
$N, \Lambda, T_a, k_0$. For simplicity we use the same $T_a$ for all
channels.

\begin{figure}
  \begin{center}
  \resizebox{\columnwidth}{!}{\includegraphics{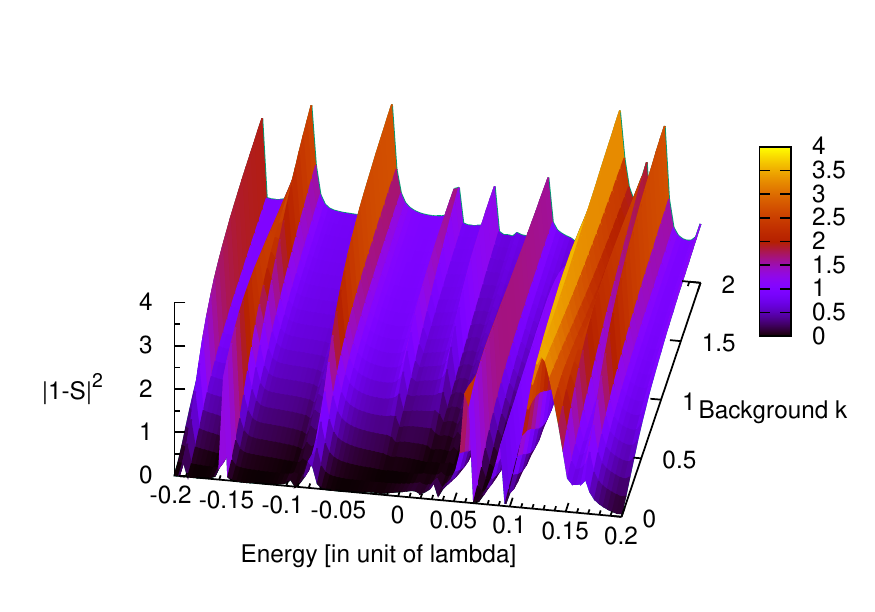}}\\
  \resizebox{\columnwidth}{!}{\includegraphics{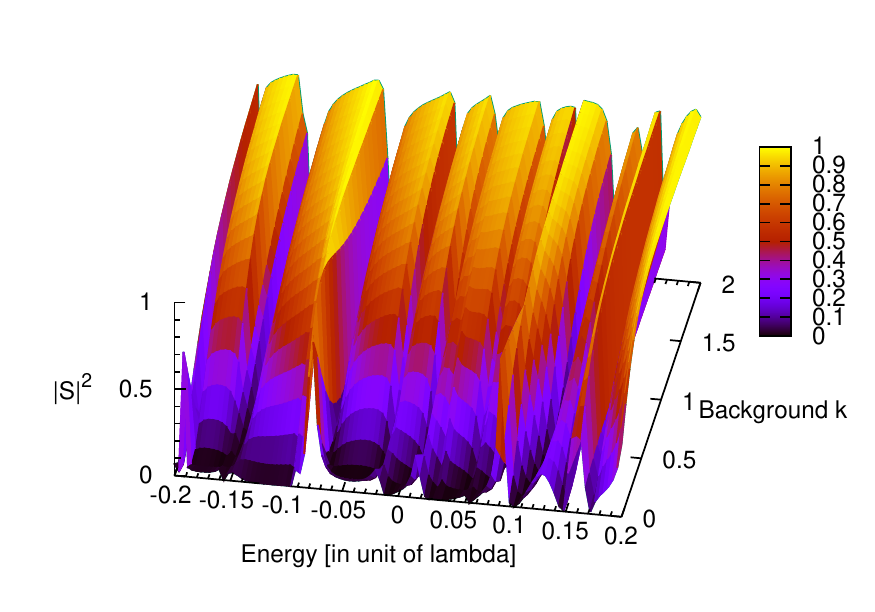}}
  \end{center}
  \caption{(Color Online) Simulated elastic (top panel) and inelastic
  (bottom panel) scattering cross sections as functions of the
  background parameter $k_0$.}
  \label{fig:cdir}
\end{figure}

The cross sections are calculated in the following three ways.
\begin{itemize}
\item For each value of $k_0$, the MC method is used to generate 
100,000 realizations of $S$. The average cross section is obtained
directly as the average of $|\delta_{ab}-S_{ab}|^2$ over that ensemble. 
Figure~\ref{fig:cdir} shows the cross sections for $N=100$, $\Lambda =
2$, $T_a=0.8 = T_b$, and for $k_0$ varying from 0 to 2 obtained in
that way. The top panel shows the elastic scattering cross section $|1
- S_{aa}|^2$, the bottom panel shows the inelastic scattering cross
section $|S_{ab}|^2$.

\item The average of $S$ over the ensemble of 100,000 realizations is
used to calculate the  modified transmission coefficients $T'_a$ of
Eq.~(\ref{eq:Tmod}). These are used in the GOE triple-integral to
calculate the width fluctuation correction.

\item $\overline{S}$ as obtained in the previous step is diagonalized
using the EW transformation. The eigenvalues $p_a$ are used in the
GOE triple-integral. The result $\overline{ \tilde{S}_{pq}
\tilde{S}_{rs}^* }$ is back-transformed to $\overline{ |S_{ab}|^2 }$.
\end{itemize}

We analyze our results in terms of the usual ``optical
model'' cross sections
\begin{eqnarray}
\sigma_{\rm T} &=& 2(1 - \Re\overline{S}_{aa}) \ ,\\
\sigma_{\rm SE} &=& |1 - \overline{S}_{aa}|^2 \ ,\\
\sigma_{\rm DI} &=& |\overline{S}_{ab}|^2 \ .
\end{eqnarray}
Here $\sigma_{\rm T}$, $\sigma_{\rm SE}$ and $\sigma_{\rm DI}$ stand
for the total, the shape elastic, and the direct inelastic cross
section, respectively. The reaction cross section and the compound
formation cross section are defined as $\sigma_{\rm R} = \sigma_{\rm
T} - \sigma_{\rm SE}$ and $\sigma_{\rm CN} = \sigma_{\rm R}
- \sigma_{\rm DI}$, respectively. All these cross sections are given
by the coupled-channels optical model, while the compound elastic
($\sigma_{\rm CE}$) and compound inelastic ($\sigma_{\rm CI}$) cross
sections require statistical-model calculations. We do not use a
coupled-channels optical model in the present context but are able to
calculate all these cross sections directly from the MC simulation.
The parameter $k_0$ controls the strength of $\sigma_{\rm DI}$ up to a
limit defined by unitarity --- since $\sigma_{\rm T}$ and $\sigma_{\rm
SE}$ are connected by $\overline{S}_{aa}$, $\sigma_{\rm R}$ is
constrained even if $k_0$ is very large.

\begin{figure}
  \begin{center} \resizebox{\columnwidth}{!}{\includegraphics{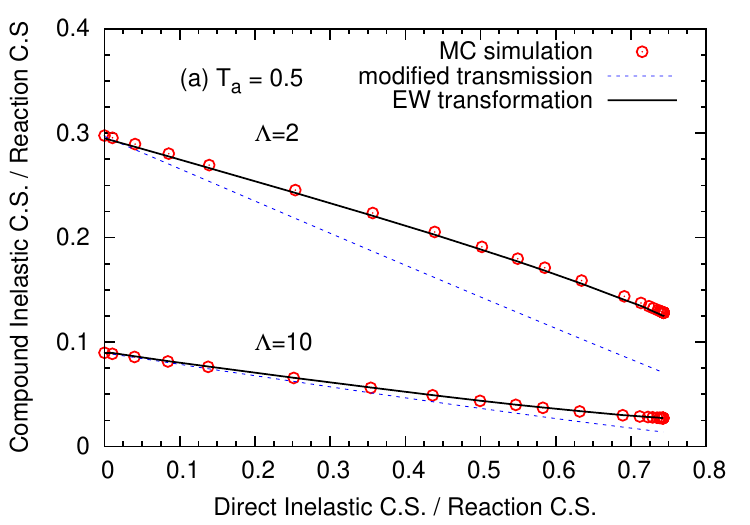}}
\\ \resizebox{\columnwidth}{!}{\includegraphics{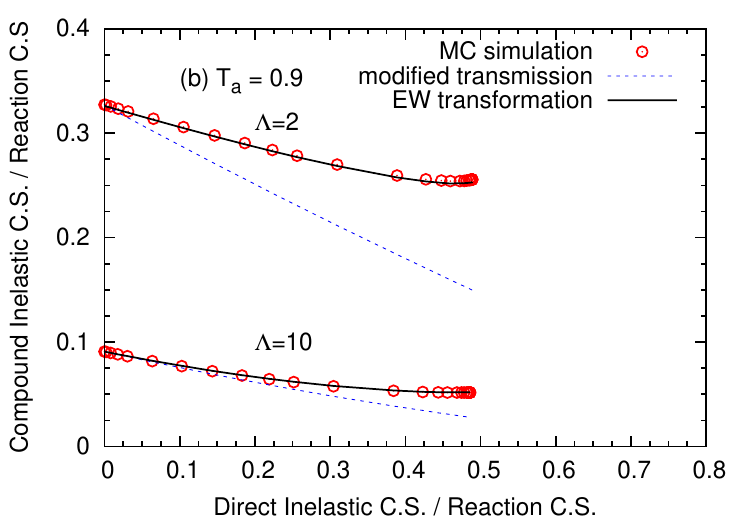}} \end{center} 
\caption{(Color
  Online) Ratio $\sigma_{\rm CI}/\sigma_{\rm R}$ of the compound
  inelastic scattering cross section to the reaction cross section as
  a function of the ratio $\sigma_{\rm DI} / \sigma_{\rm R}$ of direct
  reaction cross section to the reaction cross section. The symbols
  show the ensemble average of the MC simulation, the dotted lines are
  cross sections calculated with the modified transmission coefficients
  $T'_a$, and the solid lines are the result of the EW transformation.
  The top panel is for $T_a=0.5$, and the bottom panel is for
  $T_a=0.9$.}  \label{fig:crxewt}
\end{figure}

Figure~\ref{fig:crxewt} shows how the compound-inelastic scattering
cross section changes with the strength of the direct reaction. We
plot the ratio of $\sigma_{\rm CI}$ to the reaction cross section
$\sigma_{\rm R}$ as a function of the ratio $\sigma_{\rm DI} /
\sigma_{\rm R}$. The upper panel is for $T_a=0.5$ and the lower
panel is for $T_a=0.9$. In each panel we show two cases, $\Lambda = 2$
and 10. The results from the EW transformation agree perfectly with
the MC simulations, confirming that the EW transformation with the GOE
triple-integral yields the correct average cross section when there
are strongly coupled channels.
 
When the background $K$-matrix is parametrized as in
Eq.~(\ref{eq:Kbg}), $\sigma_{\rm CI}$ approaches the unitarity limit
for very large $k_0$. At this limit, we have $\sigma_{\rm
CE} \simeq \sigma_{\rm CI}$. The elastic enhancement disappears when a
direct channel becomes very strong. Since we employed the same
transmission coefficients for all channels, the compound elastic and
inelastic scattering cross sections are equal in that limit and given
by $\sigma_{\rm CN} / \Lambda$. Use of the modified transmission
coefficients $T_a'$ overestimates $\sigma_{\rm CE}$ and underestimates
$\sigma_{\rm CI}$. The discrepancy increases with increasing
$\sigma_{\rm DI}$.

The EW transformation is definitely required to calculate the
correct compound cross sections when $\Lambda$ is small and
$\sigma_{\rm DI}/ \sigma_{\rm R}$ is larger than about 5\%. A case in
point might be a reaction induced by neutrons of several 100 keV
impinging on an actinide. Several levels of the ground-state
rotational band will be excited by the direct inelastic scattering
process. A simple coupled-channels calculation for the 300-keV
neutron-induced reaction on $^{238}$U gives $\sigma_{\rm DI}
/ \sigma_{\rm R}$ of about 0.1.  Therefore the approximate method that
uses the modified transmission coefficients $T'_a$ is expected to
result in an underestimate of $\sigma_{\rm CI}$.

\section{Conclusion}
\label{sec:directtheory}

We have investigated the statistical properties of the scattering
matrix containing a GOE Hamiltonian in the propagator. That $S$ matrix
describes general chaotic scattering and applies to compound-nuclear
reactions at low incident energies (below the precompound regime). We
have compared results for average cross sections obtained from
Monte-Carlo (MC) simulations with those from the GOE
triple integral and from statistical models. The latter give heuristic
accounts of the width fluctuation correction. In the GOE approach, the
results depend on few parameters: the number $N$ of resonances, the
number $\Lambda$ of open channels, and the average $S$ matrix
elements. Without direct reactions, the average $S$ matrix is
diagonal, and the relevant parameters are the transmission
coefficients in the channels. When the channels are strongly coupled
and the average $S$ matrix is not diagonal, the number of parameters
is correspondingly increased. Our simulations indicate the range of
validity of the heuristic models and have led to the following
conclusions:

\begin{itemize}
\item For all parameter values studied, the numerical average of
  MC-generated cross sections coincides with the result of the GOE
  triple-integral formula~(\ref{eq:GOE3int}). Although that formula is
  derived in the limit of a large number of resonances, it gives the
  correct average even if the number of resonances is small.

\item Energy average and ensemble average agree reasonably well
  (i) for isolated resonances when the width of the Lorentzian
  averaging function is one or two orders of magnitude larger than the
  average resonance spacing and (ii) in the Ericson regime when the
  width of the Lorentzian averaging function is one or two orders of
  magnitude larger than the average total width of the resonances.
  
\item In the strong-absorption limit (Ericson regime) where $\sum_a
  T_a \gg 1$, the channel degree-of-freedom $\nu_a$ is 2, different
  from Moldauer's asymptotic value of 1.78.

\item In extreme cases where a few open channels (including the
  incident channel) have very small transmission coefficients and a
  few others have transmission coefficients close to unity, the
  elastic channel is significantly enhanced. Most of the standard
  statistical models cannot predict that enhancement. The GOE triple
  integral is the only way to produce the correct average cross
  section.

\item Direct reactions (for instance, the excitation of states of a
  rotational band due to inelastic scattering) cause the average $S$
  matrix $\overline{S}$ to acquire large off-diagonal elements. Using
  the Engelbrecht-Weidenm\"{u}ller (EW) transformation we have
  diagonalized $\overline{S}$ and evaluated the GOE triple integral in
  the diagonal channel basis. The results agree with the MC
  simulations. We find that the direct reaction increases the
  inelastic cross sections while the elastic cross section is reduced.
\end{itemize}

\section*{Acknowledgment}

T. K. and P. T. carried out this work under the auspices of the
National Nuclear Security Administration of the U.S. Department of
Energy at Los Alamos National Laboratory under Contract
No. DE-AC52-06NA25396.

\appendix*
\section{Statistical models}
\label{sec:appendix}

\subsection{HRTW}

In the HRTW approach \cite{Hofmann75,Hofmann80}, an elastic
enhancement factor $W_a$ is expressed by the channel transmission
coefficient $T_a$, and all the channel cross sections are calculated
from an effective transmission coefficient $V_a$
\begin{equation}
   \langle \sigma_{ab} \rangle =
         \frac{V_a V_b}{\sum_c V_c}
         \left\{ 1 + \delta_{ab}(W_a -1) \right\} \ ,
   \label{eqA:HRTW}
\end{equation}
where $V_c$'s are determined from the unitarity of $S$-matrix, in
another word, the flux conservation. The values of $W_a$ were derived
from the statistical $K$-matrix analysis. There are two sets of $W_a$
parameterization, namely in the original paper of
Ref.~\cite{Hofmann75}, and the updated parameters in
Ref.~\cite{Hofmann80}. We refer to the updated parameters as HRTW,
which reads
\begin{eqnarray}
  W_a &=& 1+\frac{2}{1+T_a^F}
         + 87\left( \frac{T_a-\overline{T}}{T} \right)^2
            \left( \frac{T_a}{T} \right)^5 \ , \\
  F &=& 4 \frac{\overline{T}}{T} 
            \left( 1 + \frac{T_a}{T} \right)
            \left( 1 + 3 \frac{\overline{T}}{T} \right)^{-1} \ ,
 \label{eqA:HRTWSyst}
\end{eqnarray}
where $\overline{T}$ is the average value of $T_a$, and $T$
is the sum of $T_a$ for the all open channels $T = \sum_c T_c$.

\subsection{Moldauer}

The Gaussian distribution of $\gamma_{\mu a}$ yields the Porter-Thomas
distribution of $\gamma_{\mu a}^2$ when there is only one
channel. More generally, the distribution of $\gamma_{\mu a}^2$ will
be the $\chi^2$ distribution with the channel degree-of-freedom
$\nu_a$. In these circumstances, the width fluctuation correction
factor can be evaluated numerically as \cite{Moldauer75a, Moldauer75b,
  Moldauer76}
\begin{eqnarray}
    W_{ab} &=& (1+\frac{2\delta_{ab}}{\nu_a})
   \int_0^\infty\!\!\!
   \frac{dt}{F_a(t)F_b(t) \Pi_k F_k(t)^{\nu_k/2}} ,
     \label{eqA:MoldauerW}  \\
  F_k(t) &=& 1 + \frac{2}{\nu_k} \frac{T_k}{T} t \ .
\end{eqnarray}
The integration can be performed easily by changing the variable $t$
into $z$ as
\begin{equation}
   t = \frac{z}{1-z} ,\qquad \frac{dt}{dz} = \frac{1}{(1-z)^2} \ ,
\end{equation}
where $z \rightarrow 0$ for $t=0$, and $z \rightarrow 1$ for $t
\rightarrow\infty$.

In contrast to HRTW, Moldauer's prescription gives the width
fluctuation correction factor that ensures the unitarity for all the
channels when the channel degree-of-freedom $\nu_a$ is provided.
Moldauer obtained $\nu_a$ as a function of each channel transmission
coefficient $T_a$ and the sum of them $T = \sum_c T_c$ with the MC
simulation, which reads \cite{Moldauer80}
\begin{equation}
   \nu_a = 1.78 + (T_a^{1.212} - 0.78) \exp (-0.228 T) \ .
   \label{eqA:MoldauerSyst}
\end{equation}
The channel degree-of-freedom $\nu_a$ is related to the elastic
enhancement factor $W_a = 1+2/\nu_a$.

\subsection{KKM}

The model of Kawai, Kerman, McVoy \cite{Kawai73} is very different
from the MC approach of HRTW or Moldauer. The $S$-matrix is expressed
in terms of the optical $S$-matrix background, in which the energy
average of the resonance sum part will be zero. The optical model (or
the coupled-channels optical model) yields Satchler's transmission
matrix \cite{Satchler63}, and a new hermitian matrix $X$ in channel
space is defined as
\begin{equation}
  X = -\frac{1}{2} {\rm tr} X
      + \left\{
           ({\rm tr} X/2)^2 + P
        \right\}^{1/2} \ .
  \label{eqA:Xmatrix}
\end{equation}
In the overlapping resonance limit ($\Gamma/D \gg 1$), the average
cross section is written in terms of the $X$-matrix as
\begin{equation}
  \langle \sigma_{ab} \rangle = X_{aa} X_{bb} + X_{ab} X_{ba} \ .
  \label{eqA:KKM}
\end{equation}

Since Eq.~(\ref{eqA:Xmatrix}) is a non-linear equation in $X$,
one has to solve it by an iterative procedure \cite{Kawano08}.
When $\langle S\rangle$ is diagonal (no direct channel), KKM yields
an elastic enhancement factor $W_a = 2$. In other words, KKM gives
the correct asymptotic value in the Ericson regime. That same
statement applies in the case of direct reactions. This is seen
using the EW transformation.

\subsection{GOE}

The analytical expression of the correct Hauser-Feshbach cross
section, i.e. an analytical average over the GOE resonance parameter
distributions, was given by Verbaarschot, Weidenm\"{u}ller, and
Zirnbauer \cite{Verbaarschot85}, which is the so-called
triple-integral of Eq.~(\ref{eq:GOE3int}).  The result includes the
elastic enhancement and the width fluctuation correction at the same
time, which is one of the reasons we defined the width fluctuation
correction factor by Eq.~(\ref{eq:WFC}), namely the cross section
ratio to the Hauser-Feshbach formula.

\subsection{Asymptotic expansion}

An asymptotic expansion of the GOE triple-integral formula in powers
of $1/T$ is given by \cite{Weidenmuller84}
\begin{equation}
  \langle \sigma_{ab} \rangle
  \simeq (1 + \delta_{ab}) \frac{T_aT_b}{T} A + 2 \delta_{ab} \frac{T_a^2}{T^2} B,
  \label{eqA:asympt}
\end{equation}
where
\begin{eqnarray}
  A &=& 1 + \frac{1}{T}\left(1 + \frac{2}{T}\right)
        \left\{\Sigma_2 - (T_a + T_b) \right\}  \nonumber \\
    &+& \frac{5}{T^2}\Sigma_2
        \left\{\Sigma_2 - (T_a + T_b) \right\}  \nonumber \\
    &+& \frac{4}{T^2}
        \left(T_a^2 + T_aT_b + T_b^2 - \Sigma_3\right) ,\\
  B &=& (1-T_a)
        \left\{
           1 - \frac{2}{T} (1 + 2 T_a) + \frac{3}{T}\Sigma_2
        \right\} ,
\end{eqnarray}
and
\begin{equation}
  \Sigma_2 = \frac{1}{T} \sum_c T_c^2, \qquad
  \Sigma_3 = \frac{1}{T} \sum_c T_c^3 .
\end{equation}

\subsection{Ernebjerg and Herman}

Ernebjerg and Herman \cite{Ernebjerg04} generated a quasi-random set
of transmission coefficients, and compared the simulated cross
sections with Eqs. (\ref{eqA:HRTW}), (\ref{eqA:MoldauerW}), and
(\ref{eq:GOE3int}). They obtained a new parameterization of the channel
degree-of-freedom
\begin{equation}
  \nu_a = \frac{1}{1+f(T_a) T^{g(T_a)}} \ ,
  \label{eqA:EHSyst}
\end{equation}
where
\begin{eqnarray}
  f(T_a) &=& \frac{0.177}{1 - 20.337 T_a} \ , \\
  g(T_a) &=& 1 + 3.148 T_a(1-T_a) \ .
\end{eqnarray}

\subsection{Kawano and Talou}

Similar to Ernebjerg and Herman's attempt, the GOE triple-integral
calculation can be well-approximated by putting the following channel
degree-of-freedom in Moldauer's method
\begin{equation}
  \nu_a = 2 - \frac{1}{1+f} \ , \qquad
  f = \alpha \beta_1 \frac{T_a + T}{1-T_a} \ .
 \label{eqA:LANLsyst}
\end{equation}
They obtained
\begin{eqnarray}
  \alpha   &=& 0.0288 T_a + 0.246 \ , \\
  \beta_1  &=& 1 + 2.5 T_a(1-T_a) \exp(-2T) \ .
  \label{eqA:LANLsyst1}
\end{eqnarray}

In the special case of $T < 2T_a$, a better fit can be obtained with
\begin{eqnarray}
  f  &=& 3 \alpha \beta_2 x_a \left(\frac{T-T_a}{T_a}\right)^q \ ,\\
  \beta_2  &=& 1 + 2.5 T_a(1-T_a) \exp(-4T) \ ,
  \label{eqA:LANLsyst2}
\end{eqnarray}
where $q = 0.4x_a^{0.4}$ and $x_a=T_a/(1-T_a)$.

\end{document}